\documentclass[]{acmsmall-notab}
\usepackage[utf8]{inputenc}
\usepackage[ruled]{algorithm2e}

\SetAlFnt{\small}
\SetAlCapFnt{\small}
\SetAlCapNameFnt{\small}
\SetAlCapHSkip{0pt}
\IncMargin{-\parindent}

\usepackage{url}
\usepackage[caption=false]{subfig}
\usepackage{diagbox}
\usepackage{amssymb}
\usepackage{mathtools}
\usepackage{booktabs}
\usepackage[printonlyused]{acronym}
\acrodef{PET}{privacy enhancing technology}
\acrodefplural{PET}[PETs]{privacy enhancing technologies}
\acrodef{SNP}{single nucleotide polymorphism}

\usepackage[color]{changebar}
\def\todoCtd#1{%
	#1%
	\ifx&#1&...\fi%
	\endgroup
	\cbend
	\relax
}
\usepackage{fancyhdr}
\makeatletter
\def\runningfoot{\def\@runningfoot{}}
\def\firstfoot{\def\@firstfoot{}}
\makeatother

\title{Evaluating the Strength of Genomic Privacy Metrics}
\author{ISABEL WAGNER \affil{De Montfort University}}

\begin{abstract}
The genome is a unique identifier for human individuals. The genome also contains highly sensitive information, creating a high potential for misuse of genomic data (for example, genetic discrimination).
In this paper, I investigated how genomic privacy can be measured in scenarios where an adversary aims to infer a person's genomic markers by constructing probability distributions on the values of genetic variations.
I measured the strength of privacy metrics by requiring that metrics are monotonic with increasing adversary strength and uncovered serious problems with several existing metrics currently used to measure genomic privacy. 
I provide suggestions on metric selection, interpretation, and visualization, and illustrate the work flow using a case study on Alzheimer's disease. 
\end{abstract}

\begin{document}

\maketitle

\section{Introduction}
\label{sec:intro}
In 2001, Celera, Inc was the first to sequence a full human genome at a cost of about 300 million dollars. At the time of this writing, full genome sequences can be obtained at a cost of little more than \$1,000 per genome \cite{wetterstrand_dna_2016}. This has enabled a dramatic increase in the use of genomic data in health care (e.g., personalized medicine and pharmacogenomics), research (e.g., genome-wide association studies that correlate the appearance of diseases with specific locations in the genome), and forensics (e.g., paternity tests).
Unfortunately, the wide availability of genomic data also raises important privacy concerns, because a genome sequence uniquely identifies an individual. Possible violations of genomic privacy range from the re-identification of anonymous participants in genome-wide association studies (revealing a person's disease status) to genetic discrimination (for example, denial of insurance because of genetic predisposition). Moreover, because related individuals have similar genomes, sensitive information can be inferred not only about an individual but also about her/his kin. Despite these privacy concerns, currently, there is a lack of methods to measure how private a particular genomic technology is (i.e. genomic privacy metrics). As a result, technologies that preserve genomic privacy are still in their infancy.

In this paper, we investigate the strength of genomic privacy metrics.
We consider an adversary who targets an individual and aims to infer as much of the target's genome sequence as possible. We assume that the adversary uses an inference attack to compute a probability distribution for each variation in the target genome. This is a reasonable assumption, because several inference attacks have already been described, for example exploiting linkage disequilibrium \cite{ayday_personal_2013}, exploiting information from kin genomes \cite{humbert_addressing_2013,humbert_reconciling_2014}, exploiting systematic execution of genomic tests \cite{goodrich_mastermind_2009}, and using statistics to infer whether an individual participated in a genome-wide association study \cite{homer_resolving_2008,wang_learning_2009}.

\textbf{Contributions.}
We measured the strength of 23 genomic privacy metrics in three scenarios, for adversaries of different strengths.
The key indicator of a metric's strength was defined as monotonicity, i.e. that metrics should show decreasing privacy for increasing adversary strength.
Adversary strength, in turn, was measured by how close their inferences of genomic variants were to the true value.
We tested each of the 23 metrics in three possible attack scenarios: (1) a comparative evaluation with a large number of individuals; (2) an evaluation of kin privacy considering only related individuals, and (3) an evaluation focusing on risk factors for Alzheimer's disease.


Of the metrics we tested, we found that only 7 out of 23 metrics were strong across adversary types and scenarios, and whose values have an intuitive interpretation: the adversary's success rate, the amount of information leaked, health privacy (with information surprisal or relative entropy as base metric), information surprisal, percentage incorrectly classified, relative entropy, and user-specified innocence. 
Furthermore, we find that none of the metrics we tested are sufficiently reliable when used by themselves. 
Therefore, we recommend to combine multiple strong metrics to gain insight on as many different aspects of privacy as possible.

Our systematic comparison of genomic privacy metrics enables researchers, clinicians, and policy-makers to make an informed choice about the selection of privacy metrics and privacy-enhancing technologies. 
In addition, two new visualization methods that we introduced, namely heat maps and radar plots, will further help to ensure that new \aclp{PET} are evaluated in a consistent and comparable manner.

\section{Background}

\subsection{Genomics}
Although the human genome consists of about three billion DNA base pairs,
genomes from two human individuals differ only in about 0.2--0.4\% of base pairs \cite{tishkoff_implications_2004}.
Most commonly, this genetic variation comes from differences in single bases, called single nucleotide polymorphisms (SNPs, pronounced \textit{snips}) \cite{sachidanandam_map_2001}.
In most cases, a SNP has only two variants (alleles) in the human population. 
Usually, of the two SNP alleles one is more common than the other (called the major allele, \textit{A}, and the minor allele, \textit{a}, respectively). 
Because the genome of a somatic human cell is diploid, that is, it is comprised of two sets of chromosomes -- one set inherited from the father, and the other set inherited from the mother -- each SNP is present in two copies.
Therefore, a given SNP can be encoded as 0, 1, or 2 corresponding to the combinations \textit{AA}, \textit{Aa}, and \textit{aa} \cite{humbert_addressing_2013}.
Population-wide frequencies of alleles \textit{A} and \textit{a} can be estimated from a sample of human genomes; this has been done in the 1000 Genomes project\footnote{\url{http://www.1000genomes.org/}}.
Genome-wide association studies can identify SNPs associated with diseases by comparing the incidence of SNP variations between individuals who have and do not have a particular disease.

\subsection{Privacy Metrics}
Many privacy metrics have been proposed for different domains \cite{wagner_technical_2015}. 
However, many studies have shown their shortcomings, for example 
inconsistent metrics \cite{wagner_genomic_2015}, metrics that are hard to understand \cite{diaz_does_2007}, and metrics that work only in narrow scenarios \cite{kalogridis_privacy_2010}.
This is problematic, because use of a weak privacy metric can lead to an overestimation of privacy and result in privacy violations, for example the re-identification of individuals in published health data, thus linking individuals to their medical conditions \cite{sweeney_k-anonymity_2002}. 
Privacy metrics that suffer none of these shortcomings can still be weak if used on their own because some metrics are complementary -- they measure different aspects of privacy and thus need to be used in combination to form a more complete measurement of privacy \cite{murdoch_quantifying_2014,liu_linkmirage:_2016}.
These shortcomings show that existing privacy metrics exhibit a lack of consistency, reproducibility, and wider applicability. However, it is unknown which privacy metrics, and in which application domains, produce consistently good measurements of privacy. This
can not only impede and slow down privacy research \cite{he_privacy_2015,shokri_quantifying_2011,murdoch_quantifying_2014}, but also lead to real-world privacy violations \cite{sweeney_k-anonymity_2002}, and has recently led to calls for research on privacy metrics \cite{he_privacy_2015,shokri_quantifying_2011,murdoch_quantifying_2014}.

\subsection{Genomic privacy metrics} 
Broadly, privacy metrics measure characteristics of \aclp{PET} and quantify how much privacy a technology offers \cite{claus_structuring_2006}, for example, the adversary's probability to break a user's anonymity \cite{serjantov_towards_2003}, or the maximum amount of bits of private information an adversary can infer \cite{diaz_towards_2003}. 
In the context of genomic privacy, most research applies existing privacy metrics to genomic privacy scenarios \cite{ayday_privacy-preserving_2014,humbert_addressing_2013,ayday_personal_2013,samani_quantifying_2015}.
Some researchers also propose new metrics specific to genomic privacy \cite{ayday_personal_2013,ayday_protecting_2013,humbert_addressing_2013}. 
These papers generally propose or describe one or more metrics, and then use these metrics to evaluate a \acl{PET} in a given scenario.
However, they do not evaluate the strengths of the metrics, or how they differ from other metrics.
This paper aims to address this gap.
%
The closest to our work is \cite{murdoch_quantifying_2014}, which investigates the behavior of anonymity metrics, among them entropy and some of its variations.
In previous work, we have published an initial evaluation of metrics for genomic privacy \cite{wagner_genomic_2015}.

\subsection{Requirements for genomic privacy metrics} 
Traditionally, a strong privacy metric is one that can
(1) indicate, in terms understandable to lay people, how effectively the adversary can succeed \cite{alexander_engineering_2003};
(2) show both the privacy level and the portion of data not protected \cite{bertino_survey_2008};
(3) consider accuracy, uncertainty, and correctness as three aspects of the adversary's success \cite{shokri_quantifying_2011};
and (4) indicate not only the difficulty for the adversary, but also the amount of resources he needs to succeed \cite{syverson_why_2013}.
Most of these criteria apply to specific privacy metrics, but cannot be used to compare the strengths of different metrics.

In this work, we introduce a new criterion for strong privacy metrics -- monotonicity, which requires privacy metrics to show decreasing privacy for increasing adversary strength (Section \ref{sec:extended-evaluation}).
Because monotonicity can be quantified, we believe that it can be used to compare the strenghts of privacy metrics.
Furthermore, we rate understandability based on the results of our case study (Section \ref{sec:alzheimer}).

%

\section{Privacy Metrics}
\label{sec:metrics}

From our previous survey of privacy metrics \cite{wagner_technical_2015}, we selected 23 metrics that were applicable to our genomic privacy scenario.
The metrics are summarized in Table \ref{tab:privacymetrics}, and Table \ref{tab:notation} provides a reference for notation used.
Nine metrics have previously been applied in genomic privacy; the remaining metrics have been drawn from the wider privacy literature (see the \textit{Genomics Precedent} column in Table \ref{tab:privacymetrics}).
The metrics can be grouped into per-SNP metrics that compute values for each SNP separately, and per-individual metrics that compute an aggregate value for all of an individual's SNPs (see the \textit{per SNP} column).

\subsection{Excluded Metrics}

We excluded a range of privacy metrics that did not fit our assumptions. 

Differential privacy \cite{dwork_differential_2006} offers privacy guarantees for database queries. 
However, our scenario assumes that the adversary is already one step further in that he has already acquired a probability distribution on the target's SNP values. 
While differential privacy will not help evaluate privacy in our scenario, it could be used to prevent the adversary from acquiring a probability distribution in the first place.

$k$-anonymity \cite{malin_protecting_2005} states that an individual cannot be distinguished among at least $k-1$ other individuals. 
Since we assume that the adversary already knows the target individual, we know that $k=1$, and so this metric does not help us analyze privacy further.

The \textit{genomic privacy} metric introduced by \citeN{ayday_personal_2013} assumes that the adversary only aims to infer whether an individual's genome has a specific SNP or not. In contrast, in this paper we assume that the adversary aims to infer the values of SNPs, and we study only SNPs that individuals do have. 

\subsection{Included Metrics}

We group our description of included metrics by the output they measure, according to the taxonomy proposed in \cite{wagner_technical_2015}.

\begin{table}[!t]
\tbl{Notation\label{tab:notation}}{
\centering
\begin{tabular}{ll}
\toprule
$k \in \{0,1,2\}$ & Possible SNP values \\
$x_i$ & Estimated value of SNP $i$ \\
$y_i$ & True value of SNP $i$ \\
$p(x_i=y_i)$ & Probability to guess true value of SNP $i$ correctly \\
$p(x_i=k)$ & Adversary's estimate for the case that SNP $i$ has value $k$ \\
$r_i$ & Minor allele frequency of SNP $i$ \\
$\alpha$ & Threshold for adversary's probabilities \\
\bottomrule
\end{tabular}}
\end{table}

\addtocounter{footnote}{1}
\addtocounter{footnote}{1}
\begin{table}[!ht]
\tbl{Privacy Metrics\label{tab:privacymetrics}}{
\centering
\begin{tabular}{lp{0.45cm}p{1.05cm}p{2.8cm}lp{0.8cm}p{0.85cm}}
\toprule
Metric & per SNP & Genomics Precedent & Inputs & H/L\footnotemark[2] & Priv. Level\footnotemark[4] & Intuitive-ness\\
\midrule
Adversary's success rate & -- & $\checkmark$ & estimate, truth & L & ++ & ++ \\
Amount of information leaked & -- & $\checkmark$ & estimate, truth, $\alpha$ & L & ++ & ++ \\
Asymmetric entropy & -- & $\checkmark$ & estimate, truth, prior & H &  -- & -- \\
Asymmetric entropy (per SNP) & -- & $\checkmark$ & estimate, truth, prior & H &  o & -- \\
Coefficient of determination $r^2$ & -- & -- &  estimate, truth & L & -- & o \\
Conditional entropy & $\checkmark$ & -- & estimate, truth & H & o & -- \\
Conditional privacy loss & $\checkmark$ & -- & estimate, truth & L & o & -- \\
Cumulative entropy & -- & -- & estimate & H & o & + \\
Entropy $H(X_i)$ & $\checkmark$ & -- & estimate & H & o & + \\
Expected estimation error & $\checkmark$ & $\checkmark$ & estimate, truth & H & + & o \\
Health privacy & -- & $\checkmark$ & base metric, $c_i$ & H/L & +/++ \footnotemark[\value{footnote}] & + \footnotemark[\value{footnote}]\\
Information surprisal & $\checkmark$ & -- &  estimate, truth & H & + & + \\
Inherent privacy & $\checkmark$ & -- & estimate & H & o & --\\
Max-entropy $H_0(X_i)$ & -- & -- & estimate & H & -- & -- \\
Mean error & -- & $\checkmark$ & estimate, truth & H & ++ & o\\
Mean squared error & -- & -- & estimate, truth & H & ++ &  o \\
Min-entropy $H_{\infty}(X_i)$ & $\checkmark$ & -- & estimate & H & -- & o \\
Mutual information & $\checkmark$ & -- & estimate, truth & L & o & o \\
Normalized entropy & $\checkmark$ & $\checkmark$ & estimate & H & o & + \\
Normalized mutual inf. & $\checkmark$ & $\checkmark$ & estimate, truth & H & o & o \\
Perc. incorrectly classified & -- & -- & estimate, truth & H & ++ & ++\\ 
Relative entropy & $\checkmark$ & -- & estimate, truth & H & + & + \\
User-specified innocence & -- & -- & estimate, truth, $\alpha$ & H & ++ & ++ \\
Variation of information & $\checkmark$ & -- & estimate, truth & L & -- & o \\
\bottomrule
\multicolumn{7}{l}{\footnotemark[2]{high (H) or low (L) values indicate high privacy}}\\
\multicolumn{7}{l}{\footnotemark[\value{footnote}]{Provided a good/very good (+/++) base metric is used}}\addtocounter{footnote}{1}\\
\multicolumn{7}{l}{\footnotemark[\value{footnote}]{Privacy level $\leq 30$: --; $\in ]30,70[$: o, $\in [70,90]$: +, $> 90$: ++}}\\
\end{tabular}}
\end{table}

\subsubsection{Metrics Measuring the Adversary's Error}
The \textit{expected estimation error} quantifies the adversary's correctness by computing the expected distance between the adversary's estimate and the true value for every SNP \cite{humbert_addressing_2013}.
In the context of genomics, this distance is computed on the encoded SNP values.
Therefore, we have to ensure that the SNP encoding has a meaningful genomics interpretation.
For example, the encoding proposed by \citeN{humbert_addressing_2013} is meaningful, because the encoded value 1 (one each of major and minor allele) lies between 0 (two major alleles) and 2 (two minor alleles).
This metric may behave differently with a different encoding.
\[priv_{\text{EEE}} = \sum_{k\in\{0,1,2\}}{p(x_i=k)||k-y_i||}\]

The \textit{mean squared error} is computed as the squared difference between the true value and the adversary's estimate, averaged over all SNPs \cite{oya_dummies_2014}.
\[priv_{\text{MSE}} = \frac{1}{|\text{SNPs}|} \sum_{x_i \in \text{SNPs}} \{ \|x_i-y_i\|^2\}\]
Other variations of the adversary's error are the \textit{mean error} \cite{samani_quantifying_2015} and the \textit{mean error with normalized distance} \cite{humbert_anonymizing_2015}.

\textit{Percentage incorrectly classified} measures how often the highest probability in the adversary's estimate does not correspond to true SNP value \cite{narayanan_-anonymizing_2009}.
\[priv_{\text{PIC}} = \frac{|\text{incorrect SNPs}|}{|\text{SNPs}|} \]

\subsubsection{Metrics Measuring the Adversary's Uncertainty}

\textit{Entropy} quantifies the amount of information contained in a random variable.
Used as a privacy metric, it indicates the adversary's uncertainty \cite{serjantov_towards_2003}.
\[priv_{\text{ENT}} = H(X_i) = - \sum_{k\in\{0,1,2\}}p(x_i=k) \log_2 p(x_i=k)\]
Entropy can be normalized to a range of $[0,1]$ by dividing it by Hartley entropy, that is, the logarithm of the number of outcomes \cite{humbert_addressing_2013}.
\[priv_{\text{NE}} = \frac{H(X_i)}{H_{0}(X_i)}\]

\textit{Hartley entropy}, or max-entropy, has also been used as a privacy metric \cite{claus_structuring_2006}.
It is an optimistic metric because it only accounts for the number of outcomes, but not for additional information the adversary may have.
In the context of genomics, however, the number of outcomes per SNP is known to be $3$, and therefore max-entropy is not useful and has been excluded from the evaluation.
\[priv_{\text{MXE}} = H_0(X_i) = \log_2 |x_i| = \log_2 3\]

\textit{Min-entropy} is a pessimistic metric because it is based only on the probability of the most likely outcome, regardless of whether this is also the true outcome \cite{claus_structuring_2006}.
Min-entropy is a conservative measure of how certain the adversary is of his estimate.
\[priv_{\text{MNE}} = H_\infty(X_i) = -\log_2 \max p(x_i)\]

\textit{Cumulative entropy} is based on the notion that the adversary's uncertainty increases when privacy protection is applied at several independent points. Cumulative entropy is computed as the sum of individual entropies \cite{freudiger_mix-zones_2007}.
In the context of genomics, we sum over the entropies computed for each SNP.
\[priv_{\text{CE}} = \sum_{i=1}^{|\text{SNPs}|} H(X_i)\]

\textit{Conditional entropy}, or the entropy of $X$ conditioned on $Y$, measures the amount of information needed to fully describe $X$, provided that $Y$ is known \cite{diaz_does_2007}. 
For genomic privacy, $X$ can be chosen as the true SNP value and $Y$ as the adversary's estimate. 
This measures how much more information the adversary needs to find the true value.
\[priv_{\text{COE}} = H(X_i|Y_i) = H(X_i) - I(X_i;Y_i)\]

\textit{Inherent privacy} \cite{agrawal_design_2001,andersson_fundamentals_2008} and \textit{conditional privacy} \cite{andersson_fundamentals_2008} are derivations of base metrics (entropy and conditional entropy, respectively), each computed as $2^{\text{base metric}}$.
While the base metrics are interpreted as bits of information, these metrics can be interpreted as the number of binary questions an adversary has to ask to resolve his uncertainty.
\[priv_{\text{IP}} = 2^{H(X_i)}\ ,\ priv_{\text{CP}} = 2^{H(X_i|Y_i)}\]

\textit{Asymmetric entropy} is another measure for the adversary's uncertainty. 
It is tailored to genomics because it assumes that the adversary's estimate is based on population-wide minor allele frequencies, which results in a different maximum value for entropy for each SNP  \cite{ayday_protecting_2013}.
\[priv_{\text{AE}} = \sum_{i=1}^{|\text{SNPs}|} \frac{p(x_i=y_i)(1-p(x_i=y_i))}{(-2w_i+1)p(x_i=y_i)+w_i^2}\text{, where } w_i=
\begin{dcases}
	(1-r_i)^2 & \text{if } y_i=0\\
	2r_i(1-r_i) & \text{if } y_i=1\\
	r_i^2 & \text{if } y_i=2
\end{dcases}\]
Asymmetric entropy can also be used as a per-SNP metric to measure privacy for individual SNPs.

\subsubsection{Metrics Measuring Information Gain/Loss}
The \textit{amount of leaked information} \cite{wang_learning_2009,ayday_privacy-preserving_2014} counts the number of leaked SNPs. 
A SNP is considered leaked when the adversary's estimate for the true outcome is above the threshold $\alpha$. 
A threshold of $1$ means that a SNP is considered leaked only if the adversary is absolutely certain.
Many scenarios will adopt a more conservative threshold to cover situations when the adversary is reasonably, but not absolutely, certain.
\[priv_{\text{ALI}} = |u| \text{ so that }\forall u_i \in \text{SNPs}: p(u_i=y_i) > \alpha\]

\textit{Information surprisal}, or self-information, quantifies how much information is contained in a specific outcome of a random variable \cite{chen_how_2013}.
In the context of genomics, the outcome is the true value of a SNP, and the information content is the probability the adversary assigns to this outcome.
Informally, information surprisal quantifies how surprised the adversary would be upon learning the true value of a SNP.
\[priv_{\text{IS}} = - \log_2 p(x_i = y_i)\]


\textit{Mutual information} measures how much information is shared between two random variables $X$ and $Y$ \cite{lin_using_2002}. As before, $X$ can be chosen as the true SNP value and $Y$ as the adversary's estimate.
\[priv_{\text{MI}} = I(X_i;Y_i) = H(X_i) - H(X_i|Y_i)\]
Normalized mutual information can use either Shannon entropy~\cite{zhu_anonymity_2005} or Hartley entropy \cite{humbert_addressing_2013}. In this paper we use the latter.
\[priv_{\text{NMI}} = 1 - \frac{I(X_i;Y_i)}{H_0(X_i)}\]

\textit{Conditional privacy loss} \cite{andersson_fundamentals_2008} is derived from mutual information.
While mutual information is interpreted as the bits of information shared between the true value and the adversary's estimate, conditional privacy loss can be interpreted as the number of binary questions an adversary has to ask to arrive at the true value.
\[priv_{\text{CPL}} = 1-2^{-I(X_i;Y_i)}\]

The \textit{relative entropy}, or Kullback-Leibler divergence, between two random variables $Y$ and $X$ measures the information that is lost when $X$ is used to approximate $Y$ \cite{deng_measuring_2007}.
In the context of genomics, good choices for $Y$ and $X$ are the true value and the adversary's estimate, respectively.
This measures how many additional bits of information the adversary needs to reconstruct the true value.
\[priv_{\text{RE}} = \sum_{k\in\{0,1,2\}}p(y_i=k) \log_2 \frac{p(y_i=k)}{p(x_i=k)}\]

\textit{Variation of information} is derived from mutual information so that it fulfills the conditions for a distance metric in the mathematical sense, especially the triangle inequality \cite{meila_comparing_2007}.
It describes the distance between two random variables, chosen as the true value and the adversary's estimate.
\[priv_{\text{VI}} = H(X_i) + H(Y_i) - 2I(X_i;Y_i)\]

\subsubsection{Metrics Measuring the Adversary's Success Probability}
The \textit{adversary's success rate} captures how likely it is for the adversary to succeed.
In the context of genomics, we can define success on a per-SNP basis as the probability of correctly inferring a SNP value, and aggregate to a per-individual metric by computing the average probability for all SNPs \cite{ayday_personal_2013}.
\[priv_{\text{ASR}} = \frac{1}{|\text{SNPs}|} \sum_{i \in \text{SNPs}} p(x_i=y_i)\]

\textit{User-specified innocence} can be seen as a counterpart to the amount of leaked information, because it counts the number of SNPs that remain private \cite{chen_measuring_2012}.
A SNP is considered private if the adversary's estimate for the true outcome is below the threshold $\alpha$.
A threshold of $0$ means that a SNP is considered private only if the adversary considers it impossible.
Many scenarios will therefore adopt a higher threshold.
\[priv_{\text{USI}} = |u| \text{ so that }\forall u_i \in \text{SNPs}: p(u_i=y_i) \leq \alpha\]

\subsubsection{Metrics Measuring Similarity/Diversity}

The \textit{coefficient of determination $r^2$} describes how well a statistical model approximates data.
It is typically used for linear regression where a value of $1$ indicates a perfect fit \cite{kalogridis_privacy_2010}.
In the context of genomics, the adversary's estimate can be used as statistical model, and the true SNP values represent the data.
\[priv_{\text{R2}} = 1-\frac{SS_E}{SS_R+SS_E}\text{, where }SS_E = \sum_i (y_i-x_i)^2, SS_R = \sum_i (x_i - \bar{Y})^2\]

\subsubsection{Other Metrics}
\textit{Health privacy} focuses on those SNPs known to contribute to a specific disease.
Health privacy uses a base metric to compute per-SNP values, and then aggregates to a per-individual metric using a weighted and normalized sum \cite{humbert_addressing_2013}.
The weights $c_i$ should be chosen to reflect how much each SNP contributes to the disease.
Base metrics discussed in \cite{humbert_addressing_2013} are the expected estimation error, normalized entropy, and normalized mutual information. 
We extend this list and also investigate relative entropy, conditional entropy, information surprisal, and min-entropy as base metrics.
\[priv_{\text{HP}} = \frac{1}{\sum_{i \in S} c_i} \sum_{i \in S} c_i G_i\text{, where }G_i\text{ is a per-SNP base metric}\]

\section{Initial Evaluation}
\label{sec:evaluation}

\subsection{Data Sources}
We used two publicly available data sources for our initial evaluation.
First, we downloaded genomic data from 1857 individuals from openSNP \cite{greshake2014opensnp}. 
This dataset consists of genomic data that users acquired from 23andme\footnote{\url{https://www.23andme.com/}} and FamilyTreeDNA\footnote{\url{https://www.familytreedna.com/}} and published on openSNP.
On average, each user has data about 730k SNPs. 
This data serves as ground truth information for all metrics that rely on it (see Table \ref{tab:privacymetrics}, column \textit{Inputs}).

Second, we downloaded minor allele frequencies from the Database of Single Nucleotide Polymorphisms (dbSNP) \cite{sherry_dbsnp:_2001}.
The minor allele frequencies in this dataset are computed from a sample global population consisting of 1000 genomes.
We used minor allele frequencies to construct the \textit{reference} adversary estimate, and for the computation of asymmetric entropy.

\subsection{Adversary Models}
Our adversary models abstract from the strategies and algorithms a real-world adversary would use, and instead represent the strength of an adversary using probability distributions.
For our initial evaluation, we use two different types of adversary estimate. 
The \textit{reference} model uses the population-wide distribution of minor allele frequencies taken from the dbSNP.
Following the Hardy-Weinberg principle and denoting the minor allele frequency as $q$, the adversary assigns probabilities depending on the number of minor alleles for each SNP: for two minor alleles $p(aa)=q^2$, for two major alleles $p(AA)=(1-q)^2$, and for one each of major and minor allele $p(Aa) = 2q(1-q)$. 

The \textit{normal} model uses a series of normal distributions with a small standard deviation ($\sigma=0.1$), truncated to the $[0,1]$ range, to represent the probability that the adversary assigns to the true value. We study six strength levels with mean probabilities $\mu=0.1, 0.25, 0.4, 0.6, 0.75, 0.9$. Figure \ref{fig:cf_adversary_probs} shows the average probability the \textit{normal} adversary assigns to the true SNP value.

Intuitively, we expect that privacy is higher if the adversary's guesses are far from the true value, and lower if his guesses are close. 
For the reference estimate, we expect that the adversary's guesses are close to the true value in many cases, because the estimates are chosen to match the majority of the population.
An adversary using the reference estimate is very realistic since minor allele frequencies are easy to obtain. It is therefore important to find protection mechanisms that are effective against this kind of adversary.

\subsection{Results}
\label{subsec:evaluation}

To get a high-level overview of how the 23 metrics behave, we computed their values using all genomes in our dataset, but only 10000 SNPs each\footnote{We also evaluated the metrics using fewer genomes, but all SNPs for each. The results were very similar, which is why we report our results using the computationally much less demanding scenario with 10000 SNPs per genome.}.
We used fixed parameter values for the three metrics that have parameters. (We evaluate the effect of changing parameter settings in Section \ref{sec:parameter_settings} below.)
For health privacy, we used 1000 SNPs with equal weights, and the expected estimation error as base metric.
We set the threshold for amount of information leaked to 0.7, and for user-specified innocence to 0.3.
We computed 15 replications to make sure the results are not due to random variations in the adversary estimate, and computed confidence intervals for the mean. 
The relative errors for these confidence intervals were below 5\% in all cases, indicating that we performed enough replications to achieve highly precise results.
We implemented our computations in Python, using SciPy\footnote{\url{http://www.scipy.org/}} for the entropy-based metrics, and scikit-learn\footnote{\url{http://scikit-learn.org}} for metrics based on mutual information.

\begin{figure*}%
    \centering
	\subfloat[Expected Estimation Error]{\label{fig:as_expected_error}{\includegraphics[width=.32\linewidth]{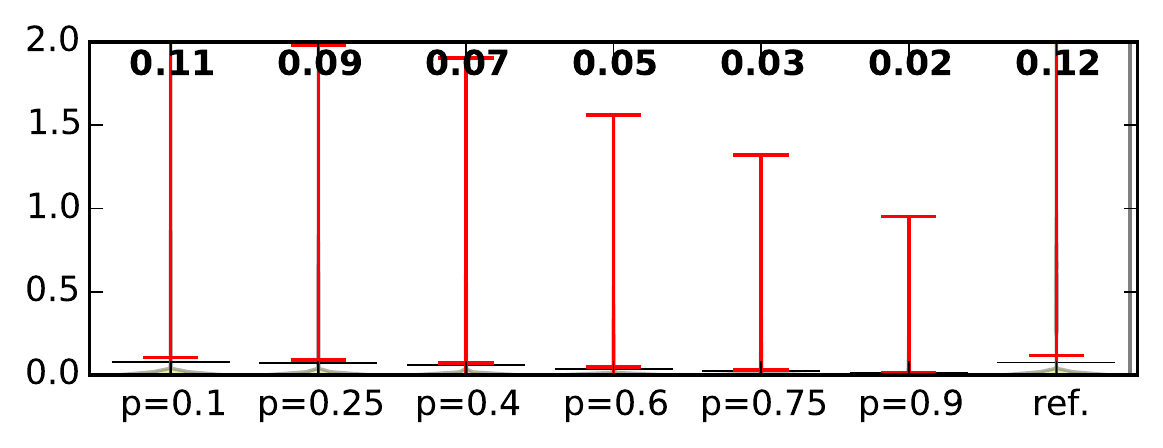} }}%
	\subfloat[Health Privacy (base: expected estimation error)]{\label{fig:as_health_privacy}{\includegraphics[width=.32\linewidth]{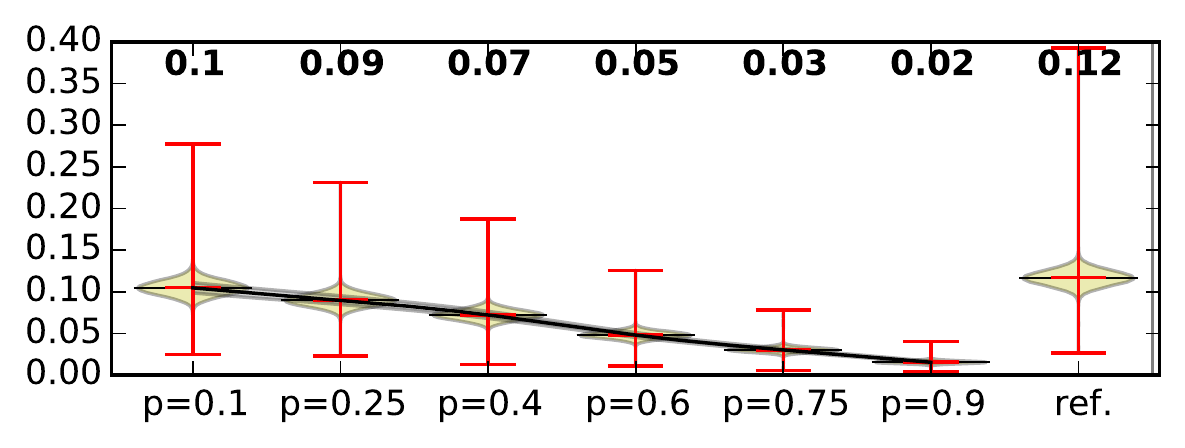} }}%
	\subfloat[Mean Error]{\label{fig:as_mean_error}{\includegraphics[width=.32\linewidth]{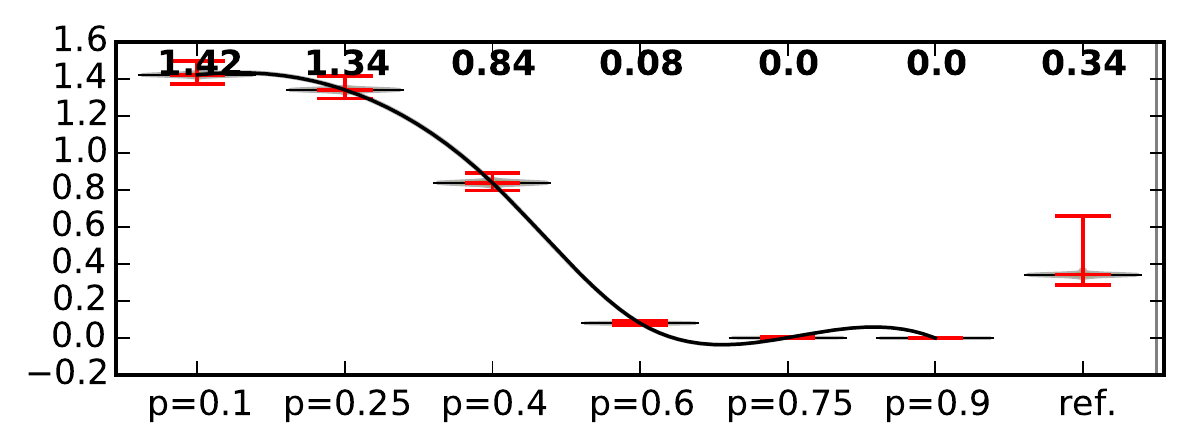} }}\\%
	\vspace{-.2cm}
	\subfloat[Mean Squared Error]{\label{fig:as_mean_squared_error}{\includegraphics[width=.32\linewidth]{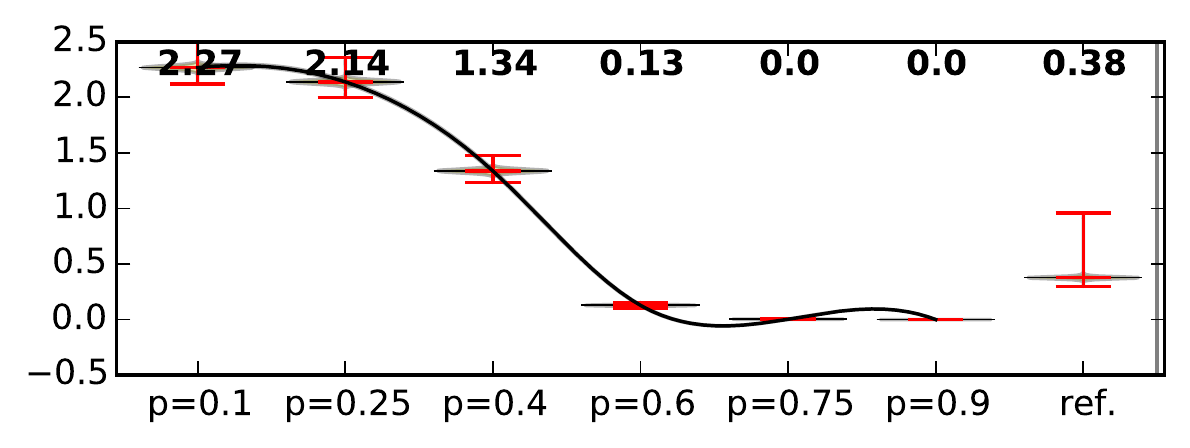} }}%
	\subfloat[Perc. Incorrectly Classified]{\label{fig:as_percentage_incorrectly_classified}{\includegraphics[width=.32\linewidth]{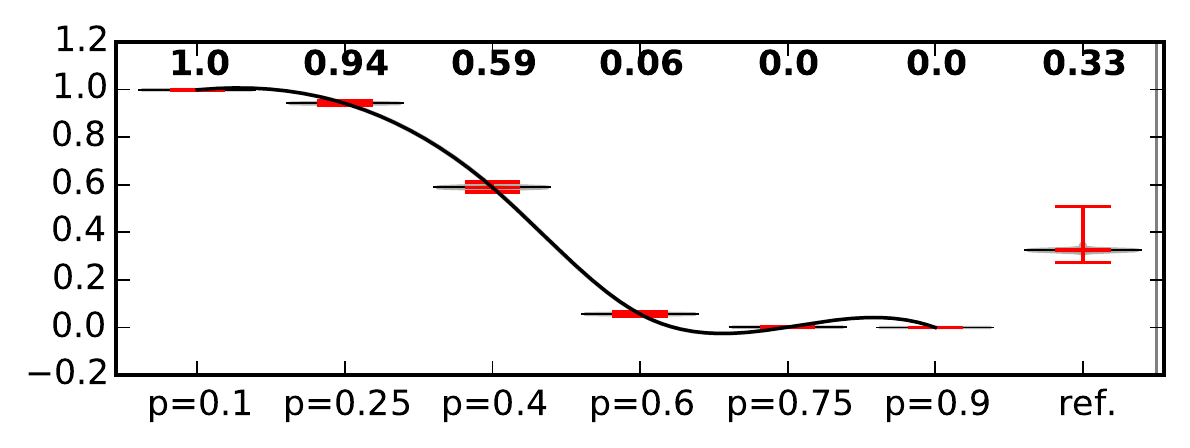} }}%
	\subfloat[Entropy]{\label{fig:as_entropy}{\includegraphics[width=.32\linewidth]{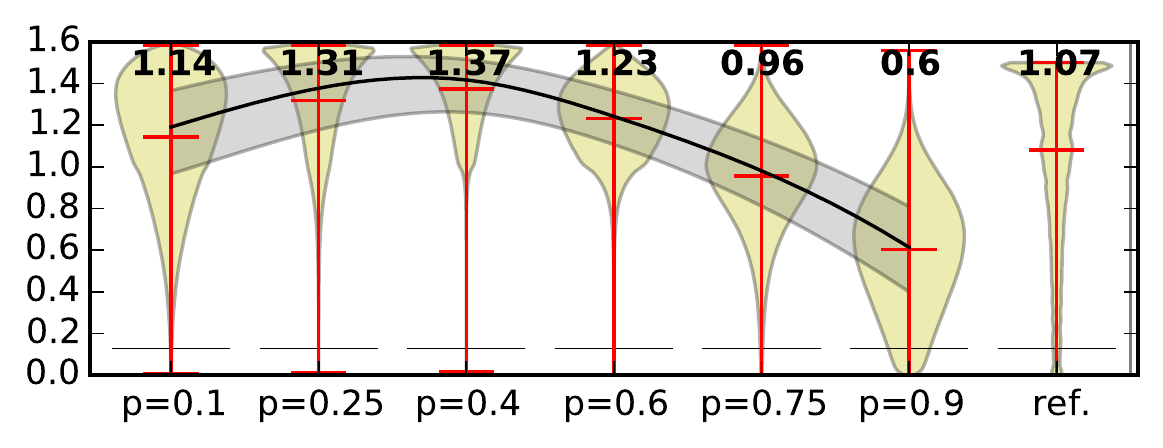} }}\\%
	\vspace{-.2cm}
	\subfloat[Asymmetric Entropy]{\label{fig:as_asymmetric_entropy}{\includegraphics[width=.32\linewidth]{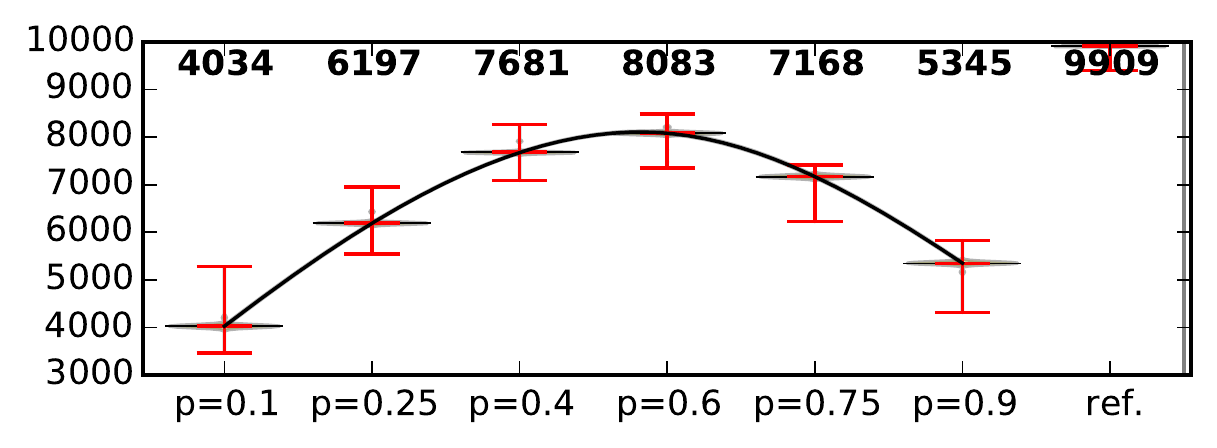} }}%
	\subfloat[Asymmetric Entropy (per SNP)]{\label{fig:as_asymmetric_entropy_per_snp}{\includegraphics[width=.32\linewidth]{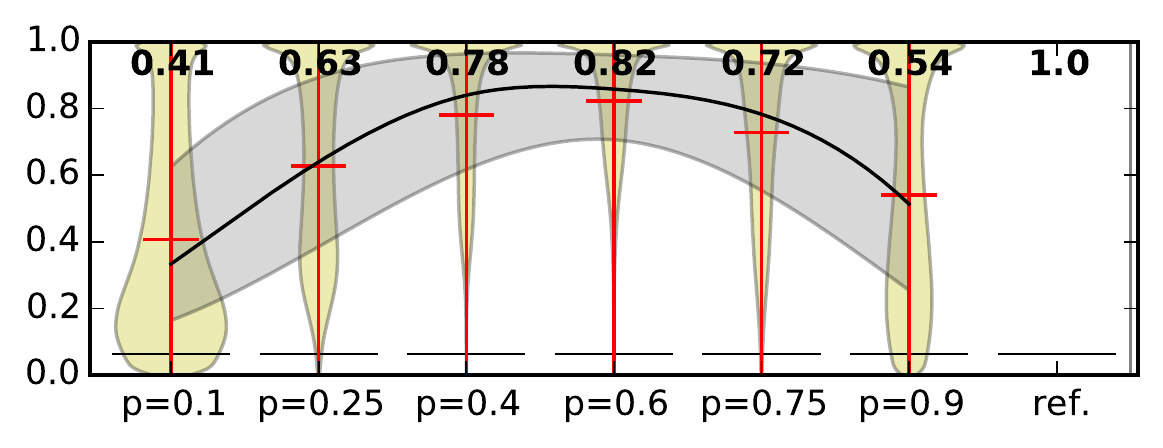} }}%
	\subfloat[Cumulative Entropy]{\label{fig:as_cumulative_entropy}{\includegraphics[width=.32\linewidth]{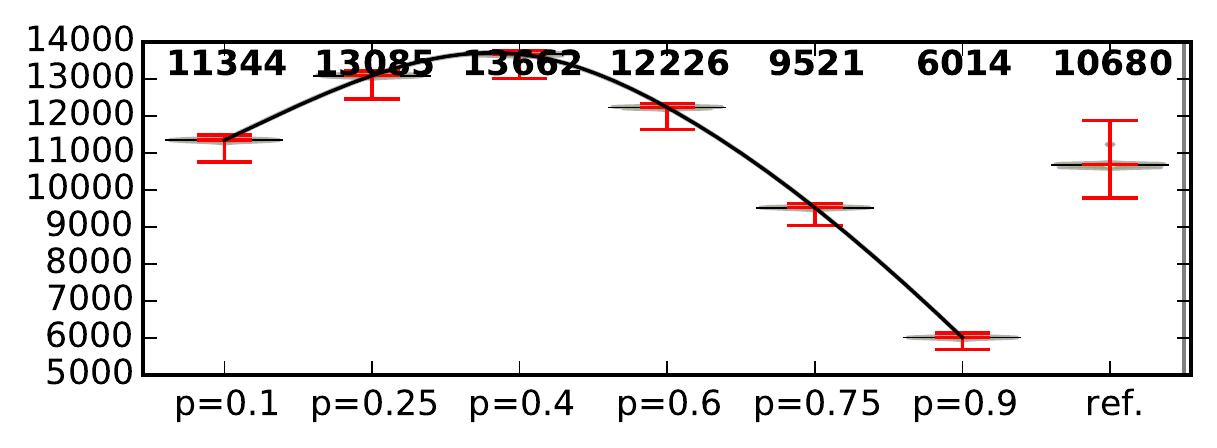} }}\\%
	\vspace{-.2cm}
	\subfloat[Normalized Entropy]{\label{fig:as_normalized_entropy}{\includegraphics[width=.32\linewidth]{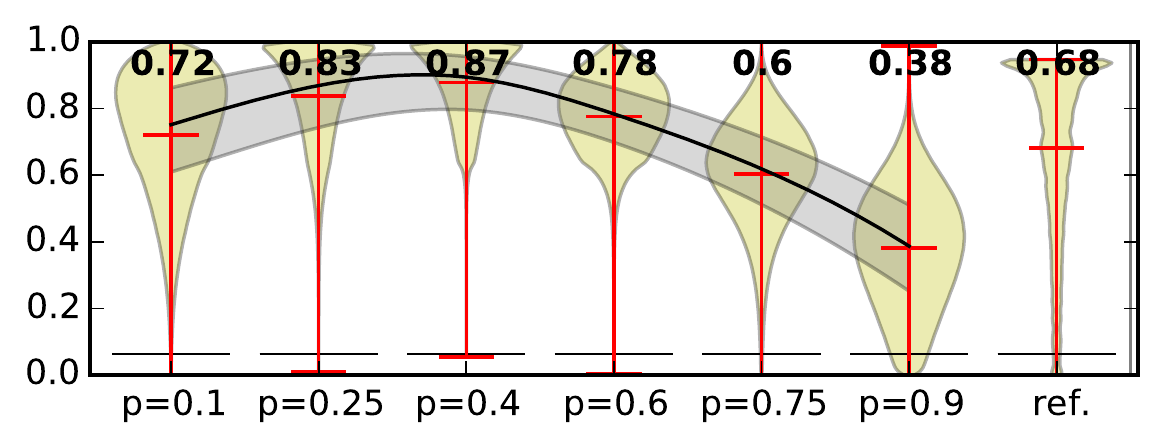} }}%
	\subfloat[Min-Entropy]{\label{fig:as_min_entropy}{\includegraphics[width=.32\linewidth]{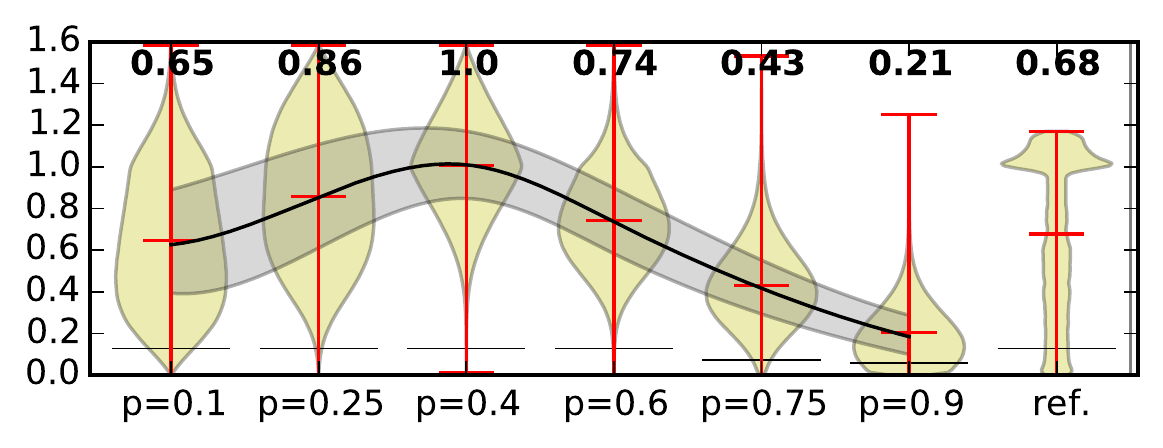} }}%
	\subfloat[Inherent Privacy]{\label{fig:as_inherent_privacy}{\includegraphics[width=.32\linewidth]{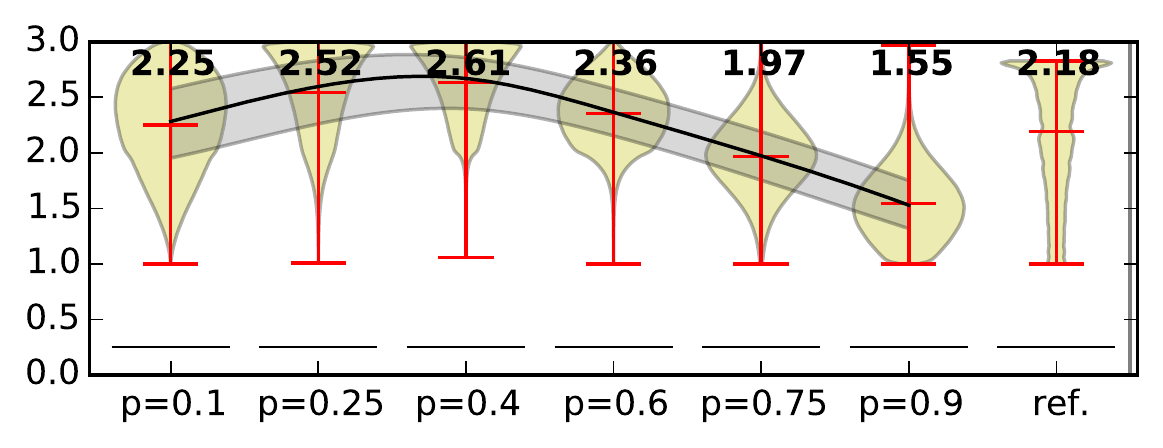} }}\\%
	\vspace{-.2cm}
	\subfloat[Conditional Entropy]{\label{fig:as_conditional_entropy}{\includegraphics[width=.32\linewidth]{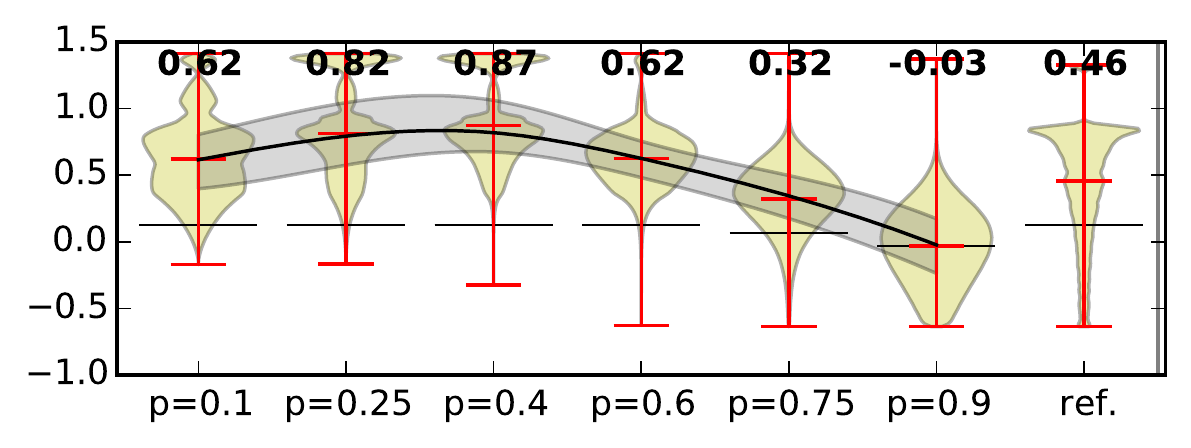} }}%
	\subfloat[Conditional Privacy Loss]{\label{fig:as_conditional_privacy_loss}{\includegraphics[width=.32\linewidth]{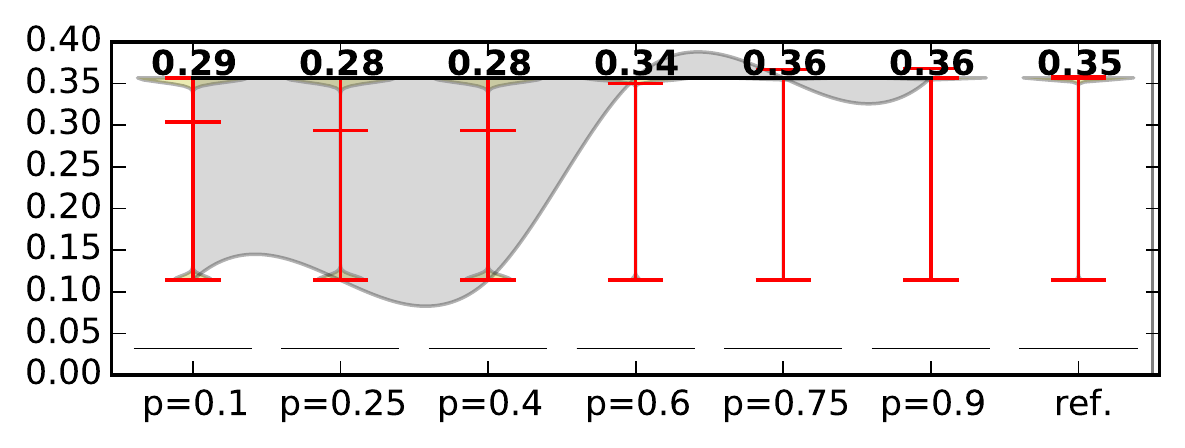} }}%
	\subfloat[Mutual Information]{\label{fig:as_mutual_information}{\includegraphics[width=.32\linewidth]{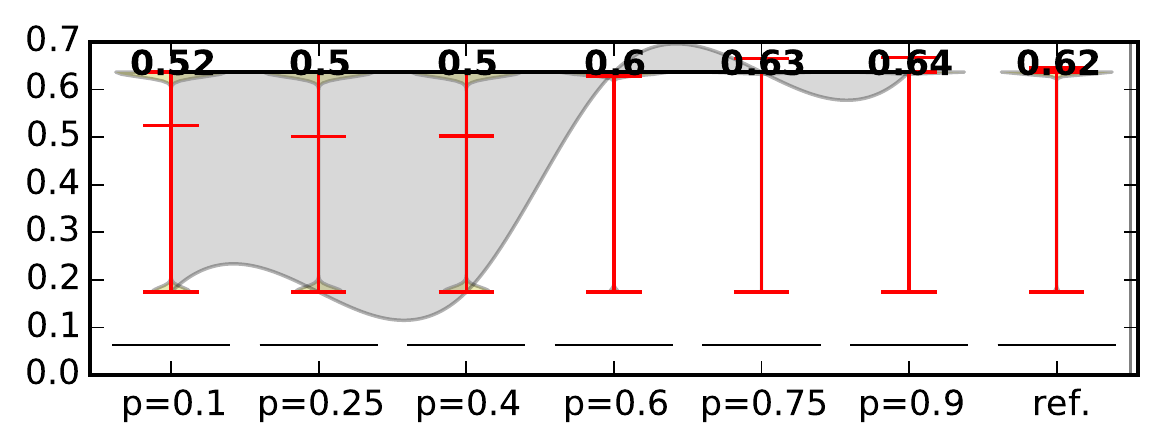} }}\\%
	\vspace{-.2cm}
	\subfloat[Norm. Mutual Information]{\label{fig:as_normalized_mutual_information}{\includegraphics[width=.32\linewidth]{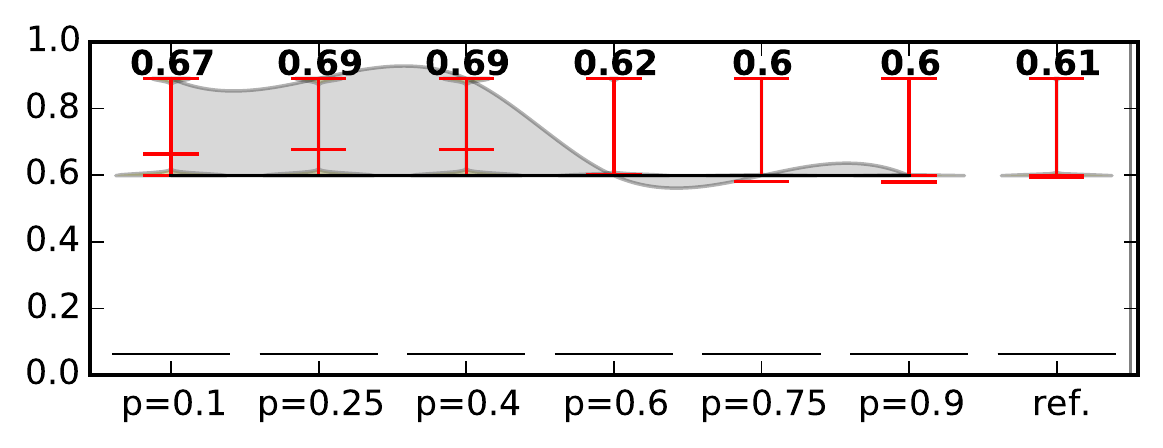} }}%
	\subfloat[Variation of Information]{\label{fig:as_variation_of_information}{\includegraphics[width=.32\linewidth]{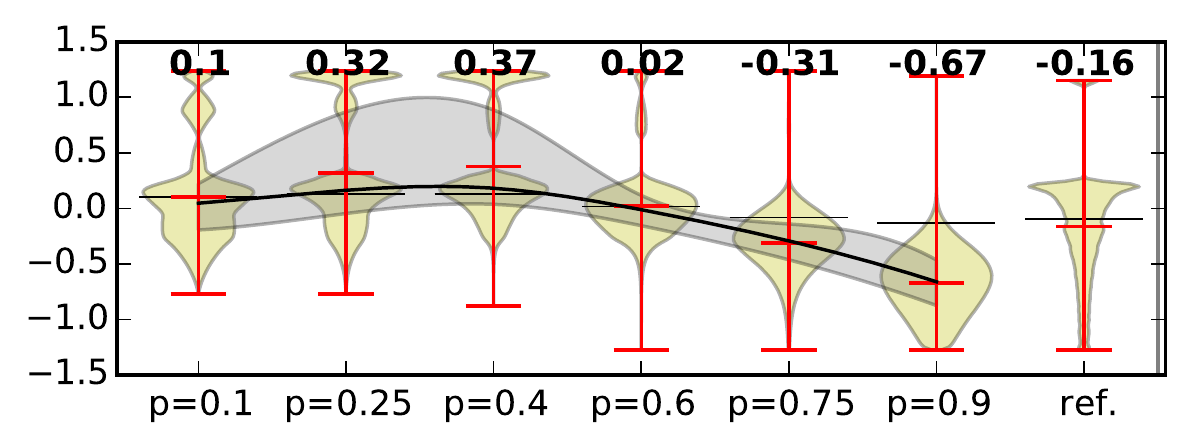} }}%
	\subfloat[Relative Entropy]{\label{fig:as_relative_entropy}{\includegraphics[width=.32\linewidth]{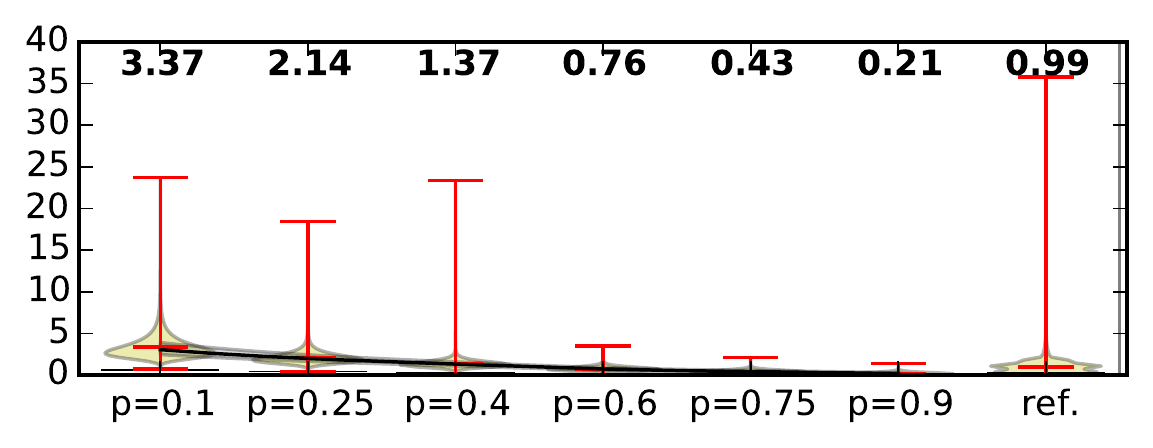} }}\\%
	\vspace{-.2cm}
	\subfloat[Information Surprisal]{\label{fig:as_information_surprisal}{\includegraphics[width=.32\linewidth]{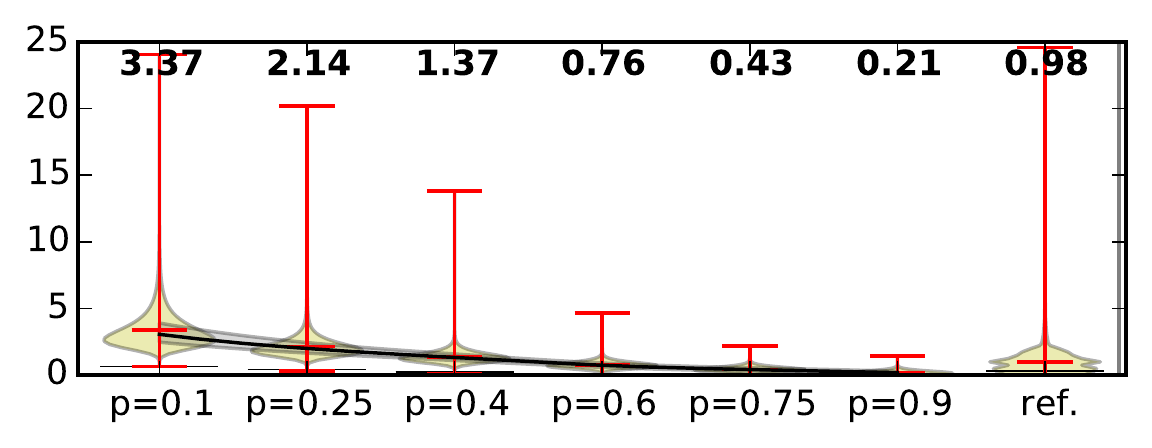} }}%
	\subfloat[Amount of Information Leaked]{\label{fig:as_information_leaked}{\includegraphics[width=.32\linewidth]{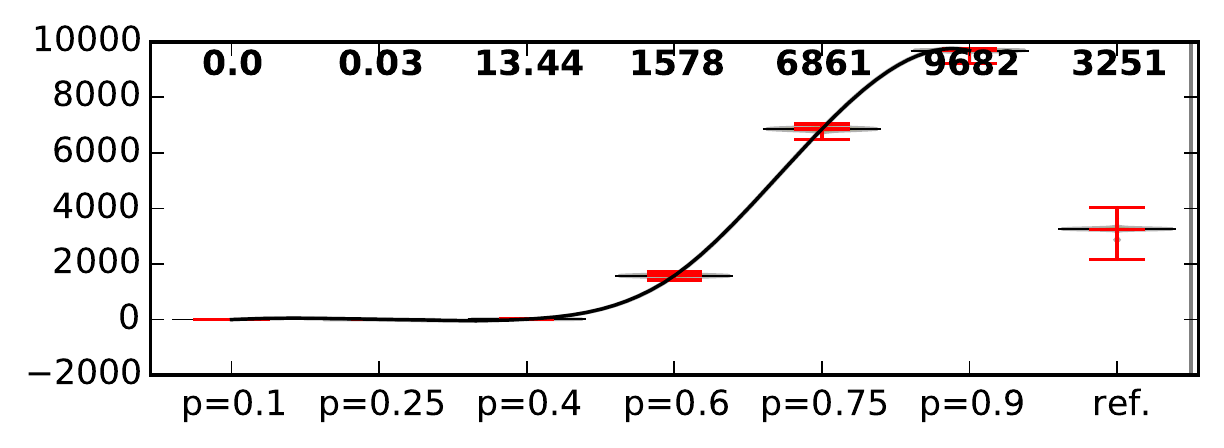} }}%
	\subfloat[User-specified Innocence]{\label{fig:as_user_specified_innocence}{\includegraphics[width=.32\linewidth]{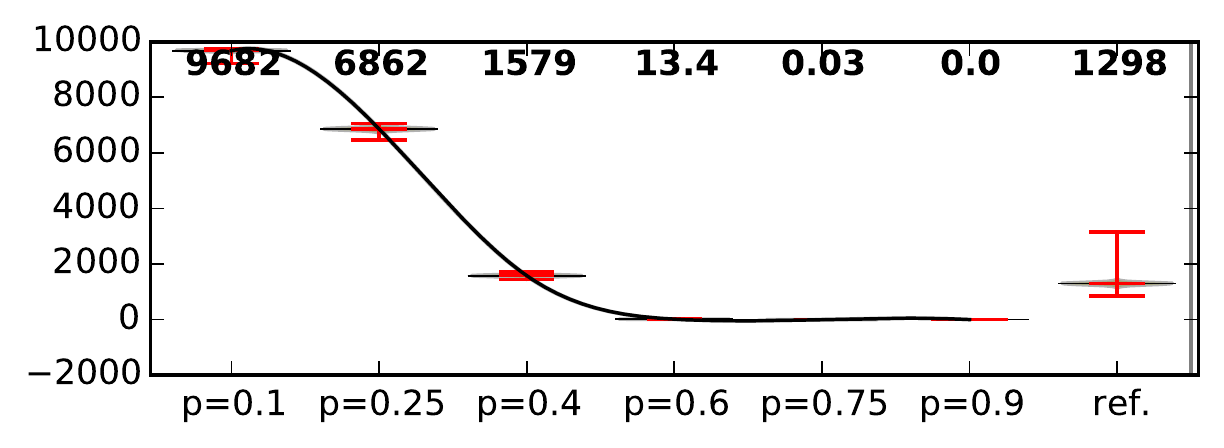} }}\\%
	\vspace{-.2cm}
	\subfloat[Adversary's Success Rate]{\label{fig:as_success_rate}{\includegraphics[width=.32\linewidth]{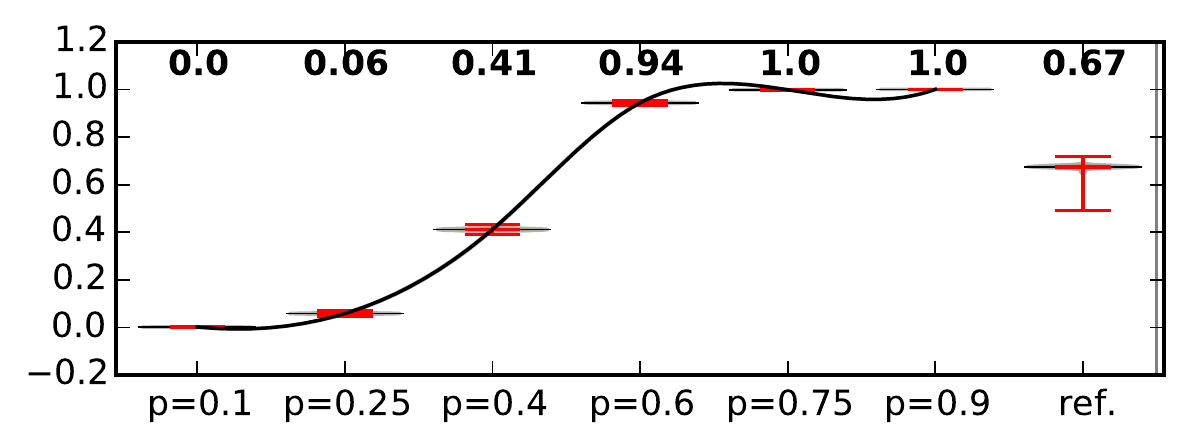} }}%
	\subfloat[Coefficient of Determination $r^2$]{\label{fig:as_coefficient_of_determination}{\includegraphics[width=.32\linewidth]{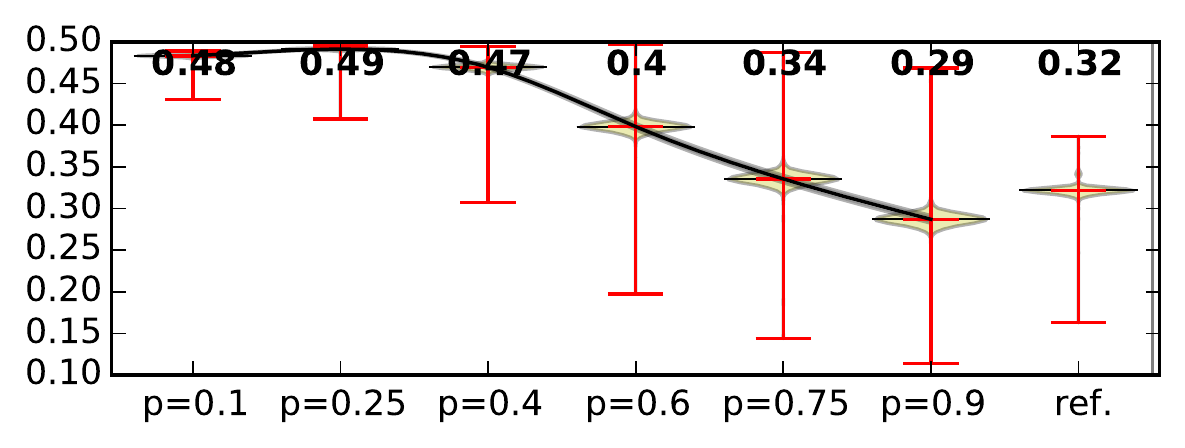} }}\\%
	\caption{Privacy metrics evaluated according to adversary strength, ordered weakest to strongest from left to right}%
    \label{fig:comparison}%
\end{figure*}

Figure \ref{fig:comparison} shows the results. For each privacy metric and adversary strength level, we plot one vertical violin plot \cite{hintze_violin_1998}.
The vertical bar shows the range of the data, with horizontal lines indicating the minimum, mean, and maximum.
In addition, a kernel density plot on each side of the bar indicates the probability density.
The six violins on the left represent the \textit{normal} adversaries with estimates ranging from far to close to the true value.
The right-most violin represents the \textit{reference} adversary.
Each of the vertical violins aggregates the results for all SNPs, all individuals, and all replications we performed for each metric and each adversary strength level.
To illustrate the statistical distribution of the bulk of metric values, we added the median as well as the first and third quartile to the plot.
We fitted cubic splines to the medians and quartiles to emphasize how their values change depending on adversary strength.
We plot the median spline as a black line, and shade the area between the quartile splines.
In addition, we print the value of the mean in boldface at the top of each violin.
We do not plot the confidence intervals since they are so narrow that the lower and upper bounds would collapse to a single line on top of the mean.


The most important requirement we look for in a privacy metric is a consistent representation of the privacy level. 
Privacy should be high for a weak adversary, and decrease with increasing adversary strength, i.e. from left to right in the plots.

\subsubsection{Metrics Measuring the Adversary's Error}
The expected estimation error (Figure \ref{fig:as_expected_error}) does not show a big difference between adversary strengths, mainly because the range of values is relatively large compared to where the bulk of the values lie. However, it can be seen that the mean is decreasing with increasing adversary strength. This becomes much more evident when the expected estimation error (a per-SNP metric) is aggregated to a per-individual metric, for example when it is used as a base metric for health privacy (Figure \ref{fig:as_health_privacy}). In this case, the decrease in value is much more pronounced (we investigate different base metrics for health privacy in Section \ref{sec:parameter_settings}).

The values for mean error, mean squared error, and percentage incorrectly classified (Figures \ref{fig:as_mean_error}, \ref{fig:as_mean_squared_error}, and \ref{fig:as_percentage_incorrectly_classified}) are decreasing, but only for the weaker four adversaries. The strongest two adversaries cannot be distinguished.

\subsubsection{Metrics Measuring the Adversary's Uncertainty}
Looking at the entropy-based metrics in Figures \ref{fig:as_entropy}--\ref{fig:as_conditional_entropy}, we can see that these metrics peak for medium-strength adversaries and assign similar values to strong and weak adversaries.
To explain this, we recall that entropy measures the uncertainty in a random variable. Because of the way we defined the \textit{normal} adversary model, medium-strength adversaries appear more uncertain than adversaries on either end.
%
While it is certainly good for privacy if the adversary is uncertain, uncertainty alone is not an accurate representation of a user's privacy level.


\subsubsection{Metrics Measuring Information Gain/Loss}
Mutual information and the metrics derived from it (conditional privacy loss and variation of information) show a similar behavior to the entropy-based metrics, albeit reversed and less pronounced (observe the horizontal bars for the mean in Figures \ref{fig:as_conditional_privacy_loss}--\ref{fig:as_variation_of_information}).

Relative entropy and information surprisal, shown in Figures \ref{fig:as_relative_entropy} and \ref{fig:as_information_surprisal}, are the only two information theoretic metrics that behave as we would expect. Their values decrease with increasing adversary strength.

The amount of information leaked (Figure \ref{fig:as_information_leaked}) and user-specified innocence (Figure \ref{fig:as_user_specified_innocence}) show the same situation from two different angles: information that is considered leaked versus information that remained private. With the parameter setting we used in this experiment, each metric can only distinguish between five of the six adversaries; the values for the weakest resp. strongest two adversaries are zero. In the other cases, the metrics behave as we expect, with increasing values for information leaked, and decreasing values for user-specified innocence. The value range for these two metrics depends on the number of SNPs in the study; the maximum value of 10000 corresponds to the number of SNPs we investigated. It would thus be easy to normalize these metrics to a range of $[0,1]$ by dividing by the number of SNPs. 
The amount of information leaked is the only metric that explicitly counts the number of information items (SNPs) not hidden by a \acl{PET}.

\subsubsection{Metrics Measuring the Adversary's Success Probability}

The adversary's success rate (Figure \ref{fig:as_success_rate}) increases with the adversary's strength, allowing to distinguish five of the six adversaries. 
The two strongest adversaries cannot be distinguished because we count a success if the estimate with the highest probability corresponds to the true value, regardless of how high this probability is.
Because we defined the adversary's success as the exact opposite of incorrect classification, the percentage incorrectly classified (Figure \ref{fig:as_percentage_incorrectly_classified}) is a mirror image of the adversary's success rate and conveys exactly the same information.

\subsubsection{Metrics Measuring Similarity/Diversity}

The values of the coefficient of determination are decreasing for most adversary strengths, as shown in Figure \ref{fig:as_coefficient_of_determination}. 
However, we expect otherwise: the lowest privacy level -- a perfect fit between the adversary's estimate and the true outcome -- should be indicated by higher values of the coefficient of determination. Figure \ref{fig:as_coefficient_of_determination} shows the reverse behavior. The coefficient of determination does therefore not give a correct estimation of a user's privacy level.

Regarding the performance of the reference adversary, we can see that most metrics place it in the middle of our adversary-strength spectrum.
Notable exceptions are the expected estimation error and health privacy, which place the reference estimate among the weakest adversaries.

\section{Extended Evaluation}
\label{sec:extended-evaluation}

We then extended our initial evaluation to address a number of open issues: 
(1) how do the genomic privacy metrics behave for datasets with different characteristics? 
(2) how do the genomic privacy metrics behave for different adversary models? 
(3) how can the results be presented in a more compact and user-friendly way? 
(4) how do the parameter settings influence the metrics' behaviors?



\subsection{Definition of Scenarios}
\label{sec:data_sources}

For the extended evaluation, we retained the two datasets from the initial evaluation (openSNP and dbSNP).
To study how relationships between individuals influence the strength of privacy metrics, we identified 13 pairs of blood relatives in the openSNP data\footnote{To identify blood relatives in the openSNP dataset, we first identified pairs of genomes that shared more than 80\% of SNP values (statistics from \url{http://www.isogg.org/wiki/Autosomal_DNA_statistics}). We then attempted to verify a potential relationship using the openSNP user names and user profiles. For 10 of these genomes, the relationship degree can be found either by references in the username (e.g., ``father of'') or by Google hits on ancestry sites, and another 3 are likely matches judging by the username (same infrequent last name), but with unknown relationship.} and added a dataset with verified blood relatives, the CEPH/Utah Pedigree 1463 \cite{drmanac_human_2010}, or \textit{Utah} for brevity, which contains the genomes of 17 family members from three generations.



We define four scenarios based on the openSNP, dbSNP, and Utah data. 
In the \textit{kin} scenario, we focus on genomes from related individuals, and evaluate all 23 privacy metrics using all SNPs for 13 pairs of related individuals from openSNP data.
In the \textit{utah} scenario, we evaluate all 23 privacy metrics using all SNPs for the 17 related individuals from the Utah data.
In the \textit{comparison} scenario, we evaluate all 23 privacy metrics using 10.000 SNPs from all genomes.
In the \textit{alzheimer} scenario, the adversary is only interested in an individual's propensity for Alzheimer's disease, and therefore we evaluate all privacy metrics on all genomes using three SNPs that are correlated with Alzheimer's disease.

\subsection{Adversary Models}
\label{sec:adversary_models}

\begin{figure}%
    \centering
    \subfloat[\textit{uniform} adversary]{\label{fig:b_adversary_probs}{\includegraphics[height=1.6cm]{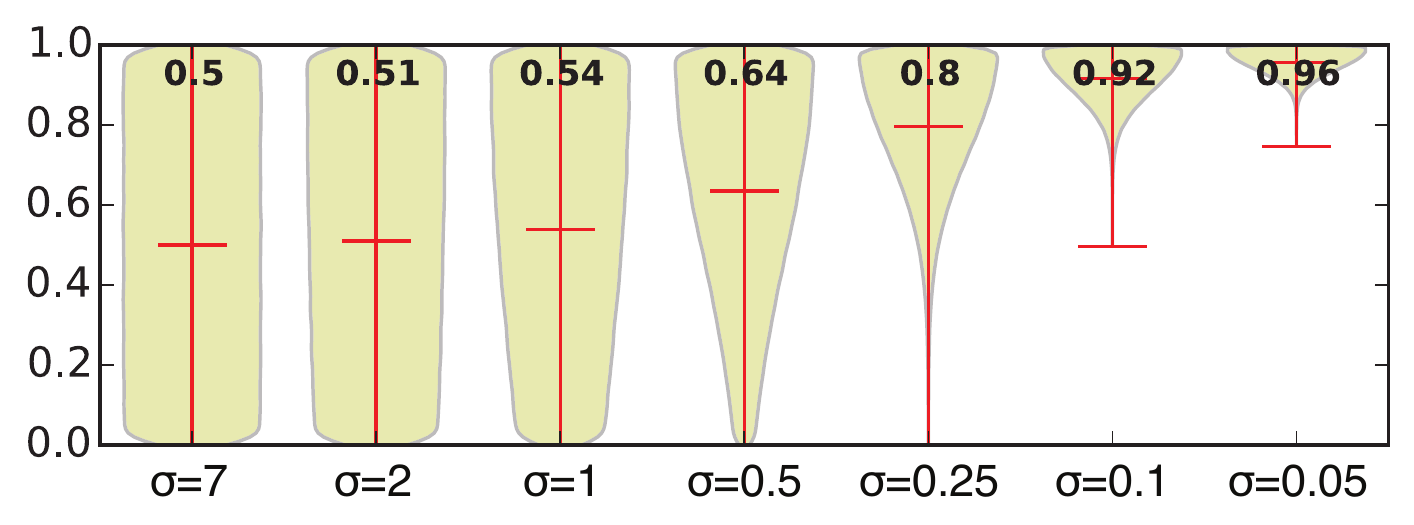} }}%
    \subfloat[\textit{normal} adversary]{\label{fig:cf_adversary_probs}{\includegraphics[height=1.6cm]{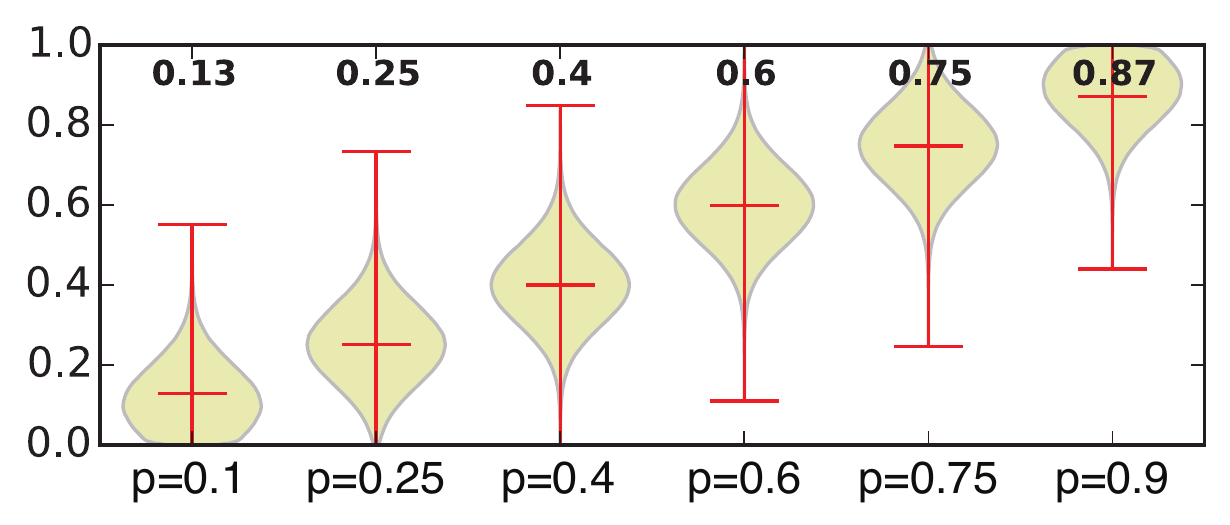} }}%
    \subfloat[\textit{reference} adversary]{\label{fig:cu_adversary_probs}{\includegraphics[height=1.6cm]{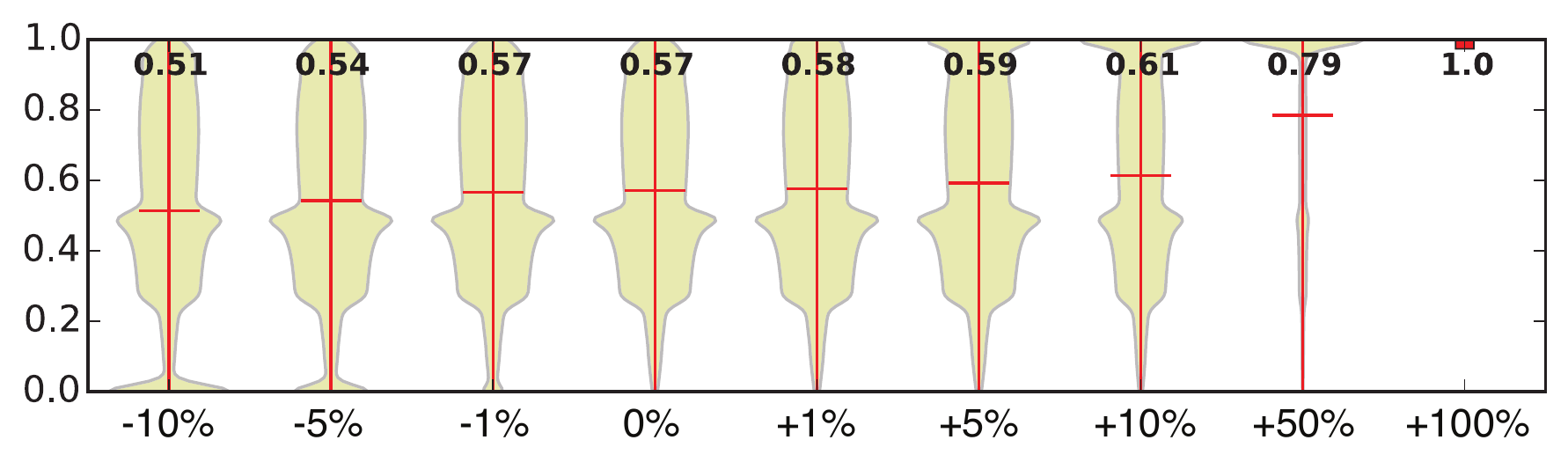} }}%
    \caption{Average probability the different adversary types assign to the true SNP value}%
\end{figure}

For the extended evaluation, we add the \textit{uniform} adversary model and extend the \textit{reference} model.
In the \textit{uniform} model, the weakest adversary makes an uninformed guess, represented by a truncated normal distribution that comes close to a uniform distribution. With increasing adversary strength, we skew the distribution towards certainty using increasingly narrow truncated normal distributions (i.e. increasingly smaller $\sigma$). Specifically, we study seven strength levels, setting the mean to $\mu=0.99$, and varying the standard deviation $\sigma=7,2,1,0.5,0.25,0.1,0.05$. Figure \ref{fig:b_adversary_probs} shows the average probability the \textit{uniform} adversary assigns to the true SNP value.

Before, we used the \textit{reference} model as an adversary with fixed strength.
Now, we vary the adversary strength by assuming that the adversary is uncertain ($p=0$) or certain ($p=1$) about some portion of SNPs. Specifically, we study nine strength levels, where the percentages indicate the portion of uncertain resp. certain SNPs: $-10\%, -5\%, -1\%, 0$ \text{(this corresponds to the fixed strength we have used before)}, $1\%, 5\%, 10\%, 50\%, 100\%$. Figure \ref{fig:cu_adversary_probs} shows the average probability the \textit{reference} adversary assigns to the true SNP value.



\subsection{Formalization of the Monotonicity Requirement}
\label{sec:monotonicity}

The initial evaluation emphasizes that a strong privacy metric should have decreasing privacy levels for increasing adversary strength.
In mathematical terms, this means that privacy metrics should be monotonic with increasing adversary strength.
Monotonicity is an important requirement, because a nonmonotonic metric can make a privacy-enhancing technology appear stronger than it actually is. If this technology is then used in practice, the use of a weak privacy metric may cause privacy violations.

We formulate an algorithm that evaluates monotonicity based on statistical tests (Algorithm \ref{alg:monotonicity}).
Our algorithm outputs heat maps as a compact and easy-to-understand visualization of the strength of privacy metrics.
Because the algorithm relies on statistics, the results are not biased by human judgment. 
In a monotonic sequence, the differences between successive pairs should all have the same sign.
The algorithm therefore awards points for each difference that has the expected sign (positive for metrics where high values indicate high privacy, and negative for metrics where low values indicate high privacy), and penalizes differences that have the wrong sign.
Because we have a large number of data points for each adversary strength level, we use statistical tests to evaluate the differences between the means of successive pairs.
Welch's t-test tests the null hypothesis that the metric values for the two adversary strengths have identical means (in contrast to the standard t-test, Welch's t-test does not assume equal variance in the two samples), and the Wilcoxon rank-sum statistic tests the null hypothesis that the metric values have been drawn from the same distribution. 
The results of these tests indicate whether the difference between the mean metric values is positive, negative, or zero, and whether the difference is statistically significant.

\begin{algorithm}[t]
\SetAlgoLined
\KwIn{arrays of metric values for each combination of adversary model and scenario}
\KwOut{heat map visualizing the strength of this privacy metric}
tests = [Welch's t-test, Wilcoxon rank-sum statistic]\\
\ForEach{combination of adversary model and scenario}{
    $m = 0$ \tcp*{holds the monotonicity points value}
    \ForEach{$test \in tests$}{
      $prevResult = 0$\tcp*{holds result for the previous pair}
      \ForEach{pair of successive adversary strengths}{
	apply $test$ to pair\\
	$p$ = statistical significance of test\\
	$result$ = value of test statistic\\
	\eIf(\tcp*[f]{test is statistically significant}){$p<0.05$} 
	{
	  \uIf{$result > 0$ ($< 0$ for lower-better metrics)}{
	      $m = m + 1$\tcp*{difference in the expected direction}
	  }
	  \uElseIf{$result < 0$ ($> 0$ for lower-better metrics)}
	    {$m = m - 1$\tcp*{difference in the wrong direction}}
	  \Else
	    {\tcp*{result is zero, do nothing}}
	}(\tcp*[f]{test is not statistically significant}){$m = m - 0.2$\\}
	\If{$result$ and $prevResult$ have different signs}
	  {$m = m - 2$\tcp*{penalize peaks in the metric value}}
	$prevResult = result$\tcp*{save result to check next pair for peaks}
      }
    }
    normalize $m$ to $[-1,1]$\\
    save $m$ for plotting
  }
plot $m$ in a heat map (rows = scenarios, columns = adversaries)
\caption{Monotonicity Computation for one Privacy Metric}
\label{alg:monotonicity}
\end{algorithm}

\begin{figure}[!ht]%
      \centering
      \includegraphics[width=\textwidth]{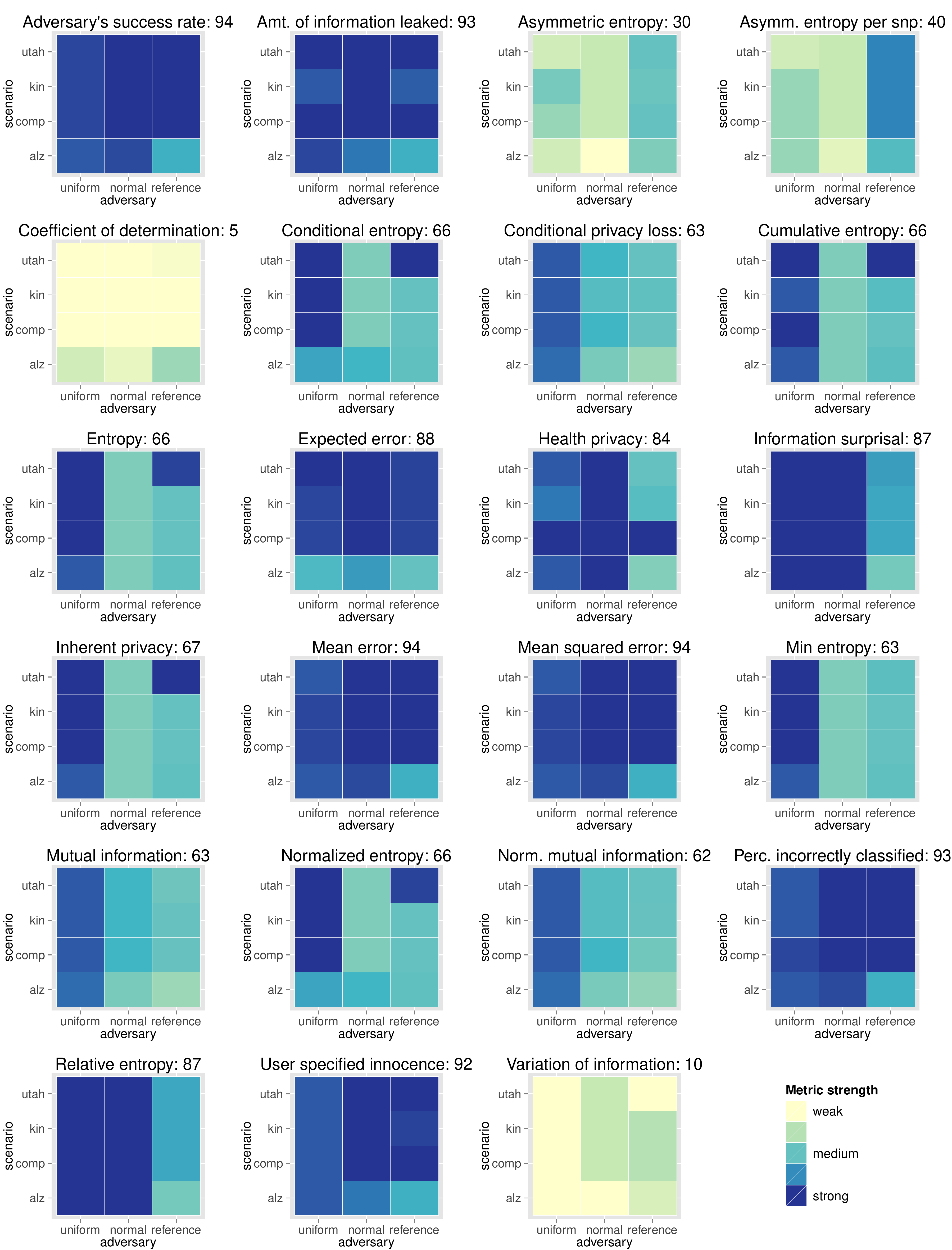}
      \caption{Strength of 23 genomic privacy metrics shown in heat maps. In each plot, the name of the metric and its overall strength are given in the title, the X axis shows the adversary model, the Y axis shows the scenario, and the colors indicate the strength of the metric (from blue=strong to yellow=weak)}%
      \label{fig:base-heat-map}%
\end{figure}

We use heat maps to visualize the resulting point values.
Figure \ref{fig:base-heat-map} shows one heat map for each privacy metric.
The rows represent scenarios (kin/utah data, kin/opensnp data, comparison, and alzheimer) and the columns represent adversary models (uniform, normal, and reference). 
Blue colors indicate a strong metric, green indicates medium strength, and yellow indicates a weak metric.
This visualization presents the strengths and weaknesses of a large number of privacy metrics in a compact way and thus helps researchers to select strong metrics. To get a sense of the overall strength of a privacy metric, we aggregate the strengths for all elements in its heat map and show it as a single percentage next to the metric name in Figure \ref{fig:base-heat-map}.

We chose the point values in our algorithm to reflect the monotonicity requirements and to create a high contrast in the visualization, which helps to pinpoint strengths and weaknesses of each metric.
We initially assigned points based on the desired behavior of a metric: a change in the right direction (1 point) is better than no change (0 points), which in turn is better than a change in the wrong direction (-1 points).
A peak (-2 points) is undesirable because it means that weak adversaries cannot be distinguished from strong adversaries.

We then conducted a sensitivity analysis to study whether the amount of points influences the final strength value.
We analyzed a full factorial design with five values for each of the five potential parameters: change in right/wrong direction, no change in mean, presence of peak, and statistical insignificance. 
The statistical analysis, using analysis of variance (ANOVA) as well as histograms, indicated that one parameter was statistically not significant (no change in mean), and our algorithm therefore assigns no points to this parameter. All other parameters were significant at $p<0.001$. The specific point values only have a small influence on the final strength level: 
the mean strength level across all parameter settings and all metrics is within 6\% of the mean value resulting from our chosen parameter setting ($<2\%$ on average).

\subsection{Results}

Figure \ref{fig:base-heat-map} shows the heat maps for the strengths of all 23 genomic privacy metrics, obtained according to Algorithm \ref{alg:monotonicity}. 
Most entropy-based metrics (asymmetric entropy, conditional entropy, cumulative entropy, min-entropy, normalized entropy, and inherent privacy) behave similarly to entropy, resulting in similar heat maps.
These metrics have clear weaknesses for two adversary types (normal and reference) and should therefore only be used in combination with other metrics.
A similar behavior, albeit less pronounced, can also be observed for mutual information, normalized mutual information, and conditional privacy loss.

Relative entropy and information surprisal are the only two information theoretic metrics that produce consistently good results, i.e. they produce consistent measurements regardless of the adversary model and scenario. 
Other strong metrics are the adversary's success and error (expected estimation error, mean error, mean squared error, percentage incorrectly classified), as well as metrics measuring the number of SNPs that are leaked or remain private.
These metrics can be recommended for use in genomic privacy.

The metrics that performed worst in our tests are the coefficient of determination, variation of information, and asymmetric entropy.
These metrics do not produce good measurements in any scenario or for any adversary model, and can therefore not be recommended for use in genomic privacy.


\subsubsection{Results for Kin Privacy}
For the majority of genomic privacy metrics, their strength does not vary when they are applied only to related individuals (\textit{kin} and \textit{utah} scenarios, top two rows of each heat map in Figure \ref{fig:base-heat-map}) as compared to a large sample of unrelated individuals (\textit{comparison} scenario, third row).
However, some metrics, for example entropy, exhibit a striking difference between the two kin privacy datasets in the strengths indicated for the reference adversary.
To explain this, we first note that both the kin and utah scenarios consist only of a small number of individuals (13 and 17, respectively).
Second, all individuals in the utah dataset are related to each other, whereas the relationships in the openSNP dataset are between pairs or groups of three.
Because the reference adversary is based on population-wide allele frequencies, the individuals in the utah dataset (top row) would all tend to have the same deviation from these frequencies, and the deviation in this particular case makes some metrics appear stronger compared to the more random sample of individuals in the openSNP dataset (second row). 
This difference in strength occurs only for medium-strength metrics, but does not occur in metrics that are strong across all adversaries and scenarios.
This emphasizes the need to select strong privacy metrics.


\subsubsection{Results for Aggregation and Normalization}

Normalization aims to bring the metric values for different scenarios into a common value range to allow comparisons.
As the heat maps for normalized entropy and normalized mutual information in Figure \ref{fig:comparison} show, normalization does not change the strength of a privacy metric with regard to monotonicity.

\begin{figure}[!ht]
 \centering
 \includegraphics[width=\textwidth]{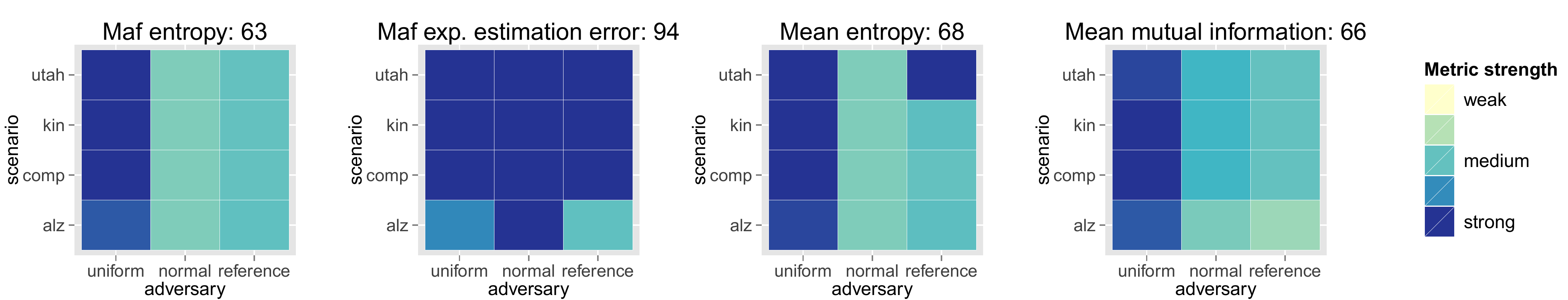}
 \caption{Strength of four aggregated privacy metrics shown in heat maps. In each plot, the name of the metric and its overall strength are given in the title, the X axis shows the adversary model, the Y axis shows the scenario, and the colors indicate the strength of the metric (from blue=strong to yellow=weak)}
 \label{fig:aggregated-metrics-heatmap}
\end{figure}

Aggregation aims to combine the metric values for all SNPs belonging to one individual, effectively reducing the amount of data to analyze.
Figure \ref{fig:aggregated-metrics-heatmap} shows the strengths of four aggregated metrics, using two different aggregation methods: an arithmetic mean and a mean weighted with population-wide minor allele frequencies (denoted \textit{maf} in the Figure).
As the comparison between the base metrics in Figure \ref{fig:base-heat-map} and the aggregated metrics in Figure \ref{fig:aggregated-metrics-heatmap} shows, aggregation does not affect the strength of privacy metrics, regardless of the aggregation method used. 

\subsection{Influence of Parameter Settings}
\label{sec:parameter_settings}

\subsubsection{Parameter Settings for Health Privacy}

Health privacy presents a way how a per-SNP metric can be aggregated into a per-individual metric.
It relies on three parameters: the selection of SNPs, the weights assigned to each, and the base metric that computes per-SNP values.
Since the metric is normalized using the sum of SNP weights, the number of SNPs and the composition of the weights do not influence the magnitude of the final value.
The value of health privacy therefore depends mostly on the value of the base metric. 

\begin{figure*}%
    \centering
    \subfloat[Normalized Entropy]{\label{fig:hp_normalized_entropy}{\includegraphics[width=\linewidth]{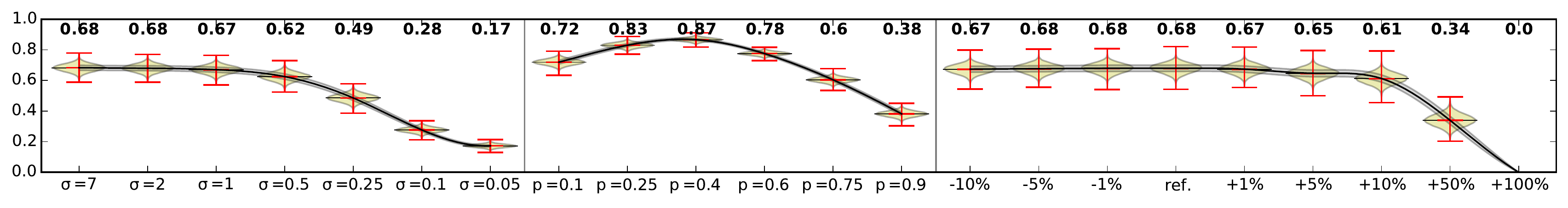} }}\\%
     \vspace{-.3cm}
    \subfloat[Normalized Mutual Information]{\label{fig:hp_normalized_mutual_information}{\includegraphics[width=\linewidth]{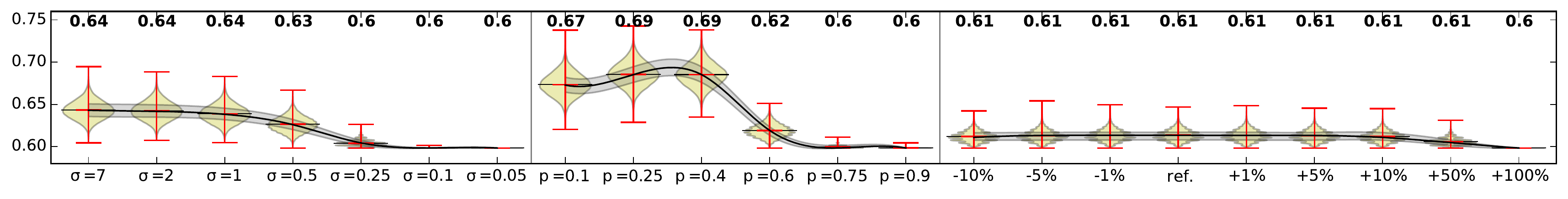} }}\\%
     \vspace{-.3cm}
    \subfloat[Expected Estimation Error]{\label{fig:hp_expected_error}{\includegraphics[width=\linewidth]{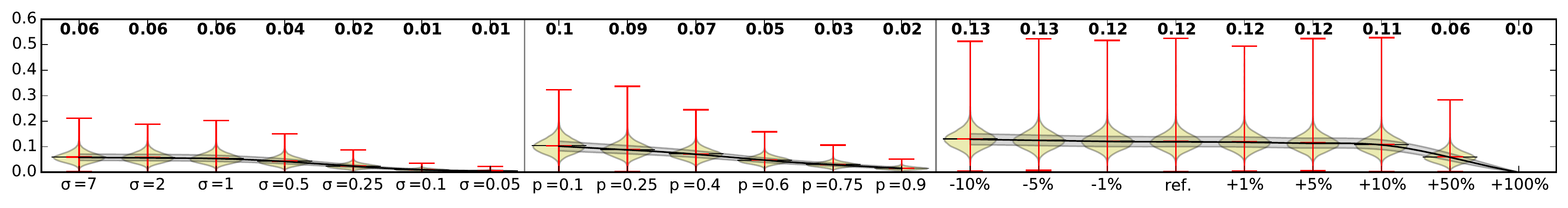} }}\\%
     \vspace{-.3cm}
    \subfloat[Information Surprisal]{\label{fig:hp_information_surprisal}{\includegraphics[width=\linewidth]{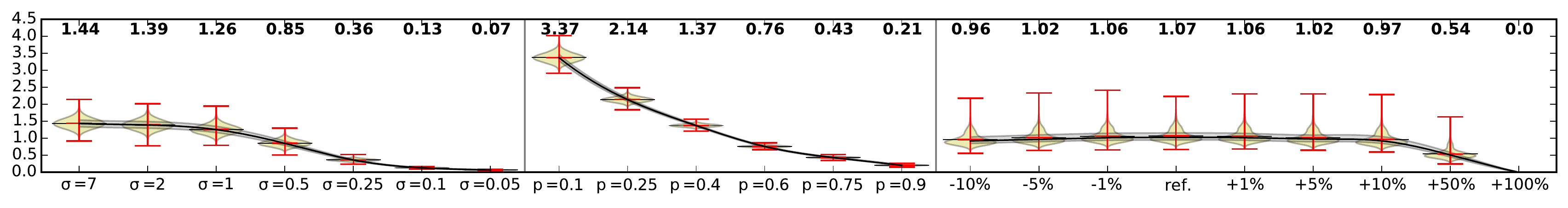} }}\\%
     \vspace{-.3cm}
    \subfloat[Relative Entropy]{\label{fig:hp_relative_entropy}{\includegraphics[width=\linewidth]{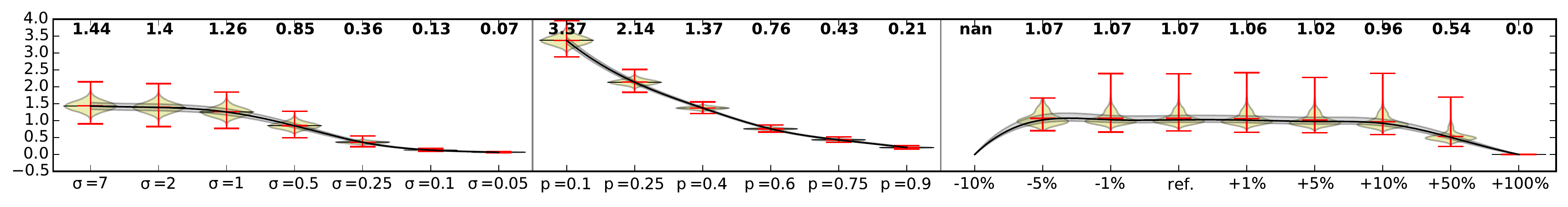} }}\\%
    \caption{Health Privacy with different base metrics, based on 100 equally weighted SNPs, evaluated using three adversary models, from left to right: \textit{uniform}, \textit{normal}, and \textit{reference}. Adversary strengths for each model are ordered weakest to strongest from left to right.}%
    \label{fig:health_privacy}%
\end{figure*}

Figure \ref{fig:health_privacy} shows health privacy for five base metrics: normalized entropy, normalized mutual information, expected estimation error, information surprisal, and relative entropy.
In every case, health privacy behaves very similar to its base metric.
All base metrics perform well for the \textit{uniform} adversary. For the \textit{normal} adversary, entropy and normalized mutual information (Figures \ref{fig:hp_normalized_entropy} and \ref{fig:hp_normalized_mutual_information}) have their highest values in the middle of the adversary spectrum and thus do not give a good indication of the achieved privacy level. Expected estimation error, relative entropy, and information surprisal (Figures \ref{fig:hp_expected_error} -- \ref{fig:hp_relative_entropy}) have strictly decreasing values for increasing adversary strengths and are thus useful to quantify the privacy level.
For the \textit{reference} adversary, most metrics are of average strength because they cannot distinguish some of the adversary strengths.

\begin{figure}[!ht]
 \centering
 \includegraphics[width=.6\textwidth]{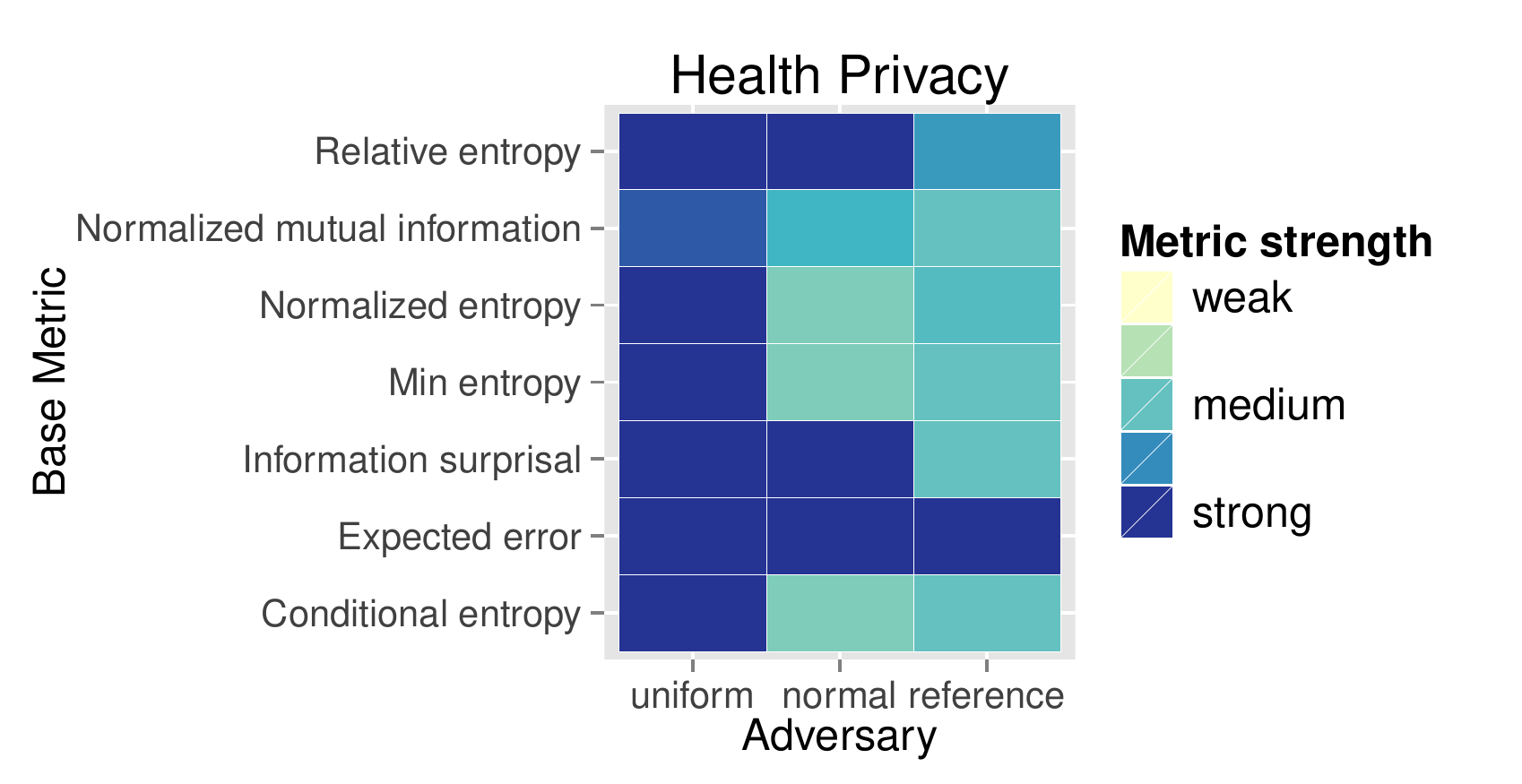}
 \caption{Strength of health privacy with different base metrics shown in a heat map. The X axis shows the adversary model, the Y axis shows the base metric, and the colors indicate the strength of the metric (from blue=strong to yellow=weak)}
 \label{fig:health-privacy-heatmap}%
\end{figure}

Figure \ref{fig:health-privacy-heatmap} shows the strengths of health privacy with different base metrics in a heat map.
We can see that the strength of health privacy corresponds to the strength of the base metric (compare with the third row in the heat maps in Figure \ref{fig:base-heat-map}).
This means that health privacy is a useful way of aggregating per-SNP metrics into a single per-individual metric, provided that the base metric is appropriate.

\subsubsection{Parameter Settings for Amount of Information Leaked and User-Specified Innocence}

Both the amount of information leaked and user-specified innocence have a threshold parameter that indicates when a SNP value is considered leaked resp. private.
We found that setting the threshold for leaked information close to $1$ resulted in zero leaked SNPs for weak adversaries, and 100\% leaked SNPs for strong adversaries.
The reverse is true for user-specified innocence, when its threshold is set to $0$.
This setting therefore doesn't allow to distinguish adversaries of different strengths.
We found that thresholds of $0.7$ for the amount of information leaked and $0.3$ for user-specified innocence allow to distinguish most of the adversary strength levels.
%
Combining the two metrics reveals additional information, because in addition to the number of leaked and private SNPs, the combination also shows for how many SNPs the leakage status is uncertain.

\section{Case study: Alzheimer's Disease}
\label{sec:alzheimer}


We applied the privacy metrics to study privacy with respect to late-onset Alzheimer's disease.
We use the case study to illustrate the process of selecting privacy metrics for a real scenario, and to show how the results can be interpreted.
This will allow us to draw further conclusions about the usefulness of metrics and metric combinations.
We identified four tasks that are necessary to measure privacy in a real genomic privacy scenario.

\subsection{Choice of SNPs for Alzheimer's Disease}

The first task is the choice of SNPs.
There are hundreds of studies correlating SNPs with Alzheimer's disease risk in the genomics literature.
We focused on three SNPs that are present in 695 genomes in the openSNP dataset: rs7412, rs429358 \cite{bertram_systematic_2007}, and rs75932628 \cite{guerreiro_trem2_2013}.
New associations with Alzheimer's are discovered frequently (see references in \cite{bertram_systematic_2007}), but because these are not available for most individuals in the openSNP dataset, we use only the three SNPs mentioned above.

\subsection{Selection of Privacy Metrics}
The second task is the selection of privacy metrics.
Following the process described in \cite{wagner_technical_2015}, we use eight questions to guide the selection:
\begin{enumerate}
 \item \textbf{Output measures.} \citeN{wagner_technical_2015} propose eight categories of output measures and suggest that metrics from as many categories as possible should be selected. Our study includes metrics from only five categories (uncertainty, information gain/loss, similarity/diversity, adversary's success probability, and error). However, the only metric belonging to the similarity/diversity category is the coefficient of determination which, as we have shown above, is not a suitable metric for genomic privacy. We will therefore select metrics from each of the other four categories.
 \item \textbf{Adversary models}. All metrics in our study are computed using the adversary's estimate. This question therefore does not influence our choice of metrics directly. 
 \item \textbf{Data source} refers to the data adversaries would use to perform their attack. In our scenario, data could be either observable or published data. Neither data source restricts our choice of metrics.
 \item \textbf{Availability of input data.} In this study, we have access to all input data required by different metrics, including knowledge of the adversary estimate, the true outcome, and parameter settings. This question does therefore not influence our choice of metrics.
 \item \textbf{Target audience.} Even though this paper is targeted at academics, some target audiences may require metrics that can be interpreted easily. We therefore discuss below how each metric can be interpreted.
 \item \textbf{Related work} in genomic privacy has used entropy, expected estimation error, adversary's success rate, and health privacy. We should therefore consider including these four metrics.
 \item \textbf{Strength of metrics.} We can refer to the heat maps in Figure \ref{fig:base-heat-map} for results about the strength of privacy metrics. The bottom row of each heat map indicates the results specific to the Alzheimer scenario we are interested in here. We list the strongest metrics in the \emph{strong metrics} column of Table \ref{tab:metric-selection}.
 \item \textbf{Implementation of metrics.} We have relied on generic implementations of entropy and mutual information available in Python packages. To the best of our knowledge, validated implementations of specific privacy metrics are not available, and therefore this question does not influence our choice of metric.
\end{enumerate}

Considering our answers to the eight questions, we see that we need to select strong metrics from four categories and include the four metrics considered in related work. Table \ref{tab:metric-selection} shows how strong metrics and metrics from related work fit into the four categories, with our choice of metrics highlighted in italics. In total, we select seven metrics: the four related work metrics, two strong metrics to add to the information gain/loss category, and one strong metric to add to the error category.

\begin{table}[!ht]
\tbl{Metric Selection\label{tab:metric-selection}}{
\centering
\begin{tabular}{lll}
\toprule
Category & Strong metrics & Related work metrics \\
\midrule
Adversary's success probability & \textit{adversary's success rate} & \textit{adversary's success rate} \\
& user-specified innocence & \\
Error & expected estimation error & \textit{expected estimation error} \\
& \textit{mean squared error} & \textit{health privacy (error)} \\
& health privacy (error) \\
& percentage incorrectly classified \\
Information gain/loss & \textit{amount of leaked information} & \\
& \textit{relative entropy} \\
& information surprisal \\
& health privacy (inf. surprisal) \\
Uncertainty &  & \textit{entropy} \\
& &  \\
\bottomrule
\end{tabular}}
\end{table}


\subsection{Choice of Metric Parameters}
The third task is to choose parameter settings for the selected metrics. 
Health privacy uses weights for individual SNPs, ideally chosen to reflect how much each SNP contributes to the overall disease risk.
This would usually be done using scientific studies or tables released by insurance companies \cite{ayday_personal_2013,humbert_addressing_2013}.
For the sake of our case study, we chose equal weights and severities for the three SNPs.

The amount of information leaked uses a parameter for the threshold probability which depends on the privacy preferences of individual users.
For our case study, we chose a threshold of 70\%, which means that SNPs are leaked if the adversary's estimate of the true value is above 70\%.


\begin{figure}
 \centering
 \subfloat[Adversary's Success Rate]{\label{fig:alz_success_rate}{\includegraphics[width=\linewidth]{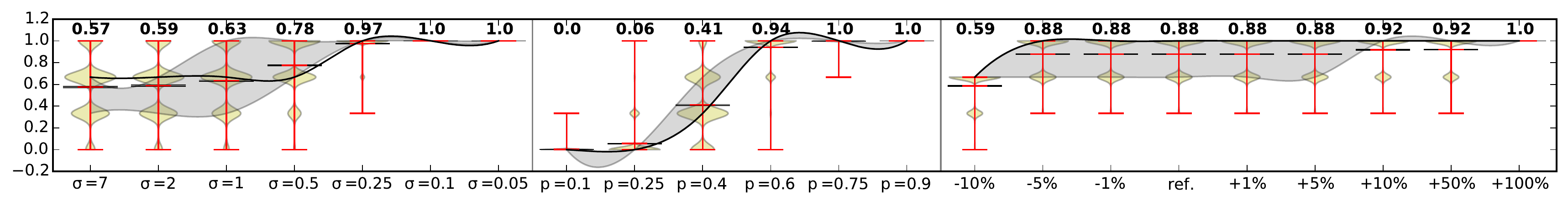} }}\\
 \vspace{-.2cm}
 \subfloat[Amount of Information Leaked]{\label{fig:alz_information_leaked}{\includegraphics[width=\linewidth]{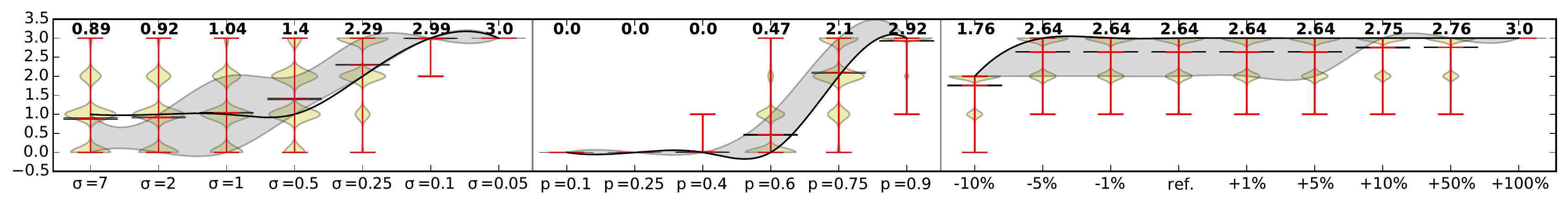} }}\\
 \vspace{-.2cm}
 \subfloat[Entropy]{\label{fig:alz_entropy}{\includegraphics[width=\linewidth]{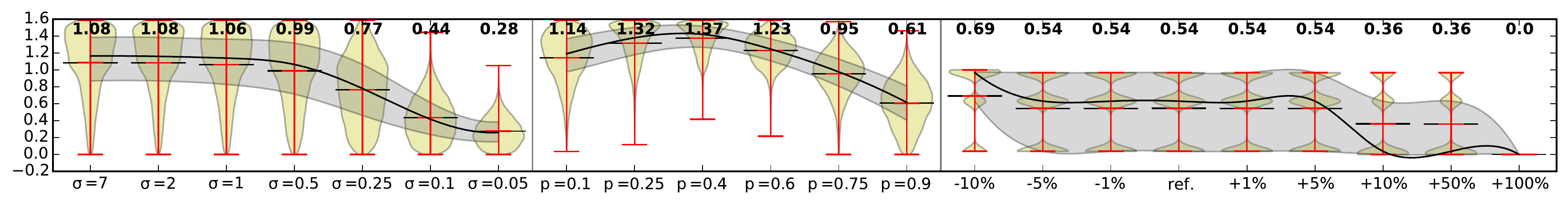} }}\\
 \vspace{-.2cm}
 \subfloat[Expected Estimation Error]{\label{fig:alz_expected_error}{\includegraphics[width=\linewidth]{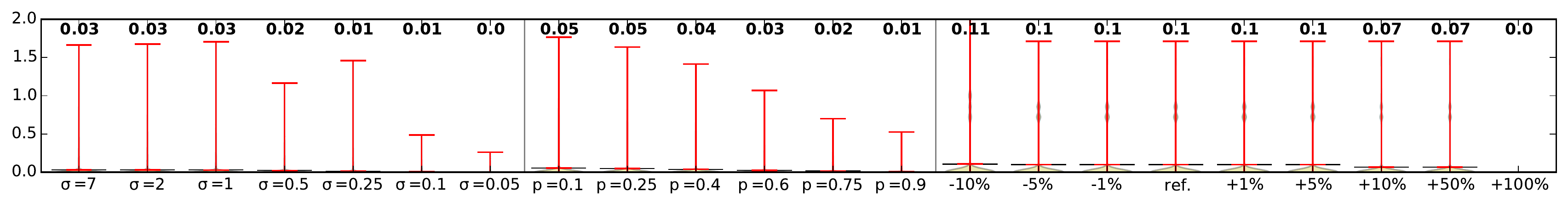} }}\\
 \vspace{-.2cm}
 \subfloat[Health Privacy (base: expected estimation error)]{\label{fig:alz_health_privacy_error}{\includegraphics[width=\linewidth]{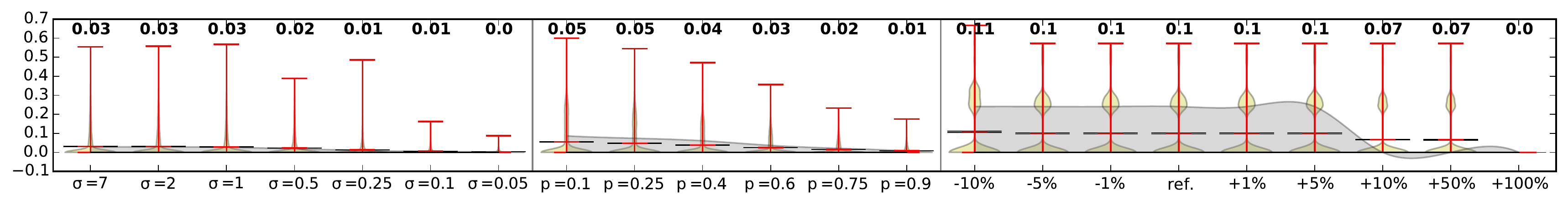} }}\\%
 \vspace{-.2cm}
 \subfloat[Mean Squared Error]{\label{fig:alz_mean_squared_error}{\includegraphics[width=\linewidth]{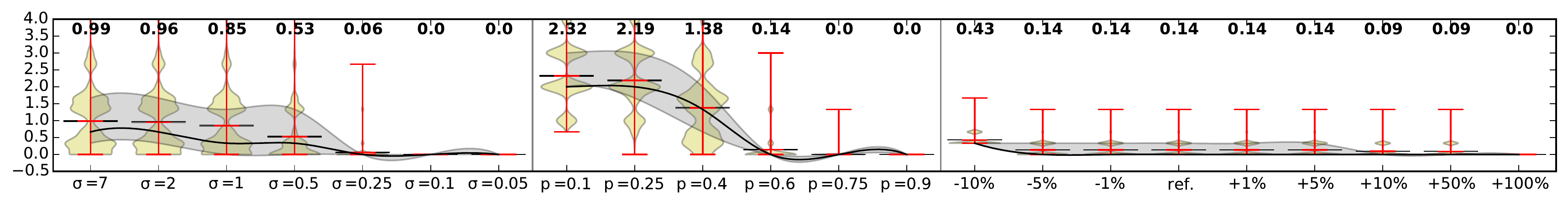} }}\\%
 \vspace{-.2cm}
 \subfloat[Relative Entropy]{\label{fig:alz_relative_entropy}{\includegraphics[width=\linewidth]{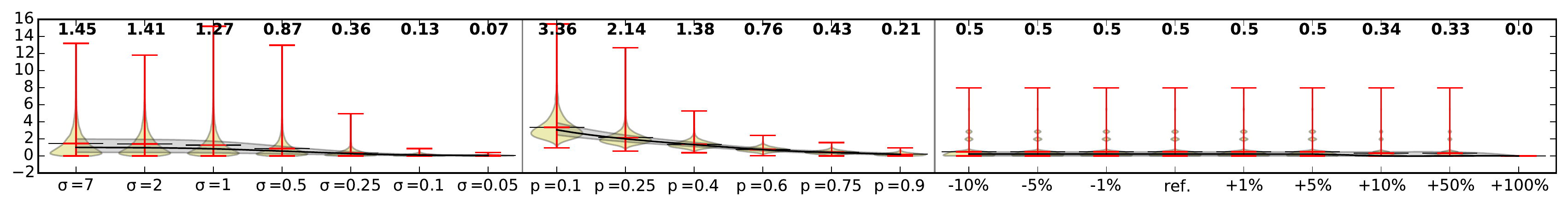} }}\\
 \caption{Strong privacy metrics for the Alzheimer's disease scenario, evaluated using three adversary models, from left to right: \textit{uniform}, \textit{normal}, and \textit{reference}. Adversary strengths for each model are ordered weakest to strongest from left to right.}%
  \label{fig:alzheimer}%
\end{figure}

\subsection{Interpretation of Results}
After conducting the privacy measurement, the fourth and final task is to plot and interpret the results.

\subsubsection{Interpreting the Values of Privacy Metrics}
To interpret what the values of each privacy metric mean, we show violin plots for the seven selected metrics in Figure \ref{fig:alzheimer}. (For completeness, we show the results for the remaining metrics in the appendix.)
For each metric, the plots show three groups of violins, corresponding to the three adversary models.
The distributions of the adversary's success rate (Figure \ref{fig:alz_success_rate}) have up to four peaks at $0, \frac{1}{3}, \frac{2}{3}$ and $1$, showing that the adversary can infer either 0, 1, 2, or 3 SNP values. 
Within each adversary group, the values are monotonic (non-decreasing), but not all strength levels can be distinguished.
%
The amount of information leaked behaves similarly to the adversary's success rate, showing how often 0, 1, 2, or 3 SNP values have leaked (Figure \ref{fig:alz_information_leaked}).
Looking at the left-most adversary group, we can see that the amount of information leaked increases at a higher adversary strength level than the adversary's success rate (compare with Figure \ref{fig:alz_success_rate}). This is because the threshold parameter for the amount of information leaked is 70\%, whereas the success rate can count successes with lower probabilities.

Entropy (Figure \ref{fig:alz_entropy}) measures the adversary's uncertainty and therefore cannot reliably indicate the user's privacy level.
Entropy performs worst for the \textit{normal} adversary model, because it peaks at a medium adversary strength.
This is consistent with the findings from Figure \ref{fig:base-heat-map}.
The values of entropy indicate how many bits of information are contained in the adversary's estimate, with high values indicating more uncertainty.

The expected estimation error (Figure \ref{fig:alz_expected_error}) and health privacy with the expected estimation error as base metric (Figure \ref{fig:alz_health_privacy_error}) behave similarly to the initial evaluation. However, both metrics show higher values for the reference adversary than for the other adversary types. This would indicate that the reference adversary performs worse (i.e., has a higher error) even than the adversary whose estimate is furthest from the true value. This differs from what the other metrics show for the reference adversary.



The mean squared error (Figure \ref{fig:alz_mean_squared_error}) shows how far, on average, the adversary's guess is from the true value. However, since the error is squared and computed on encoded SNP values, the meaning of the values is not intuitively clear.

Relative entropy (Figure \ref{fig:alz_relative_entropy}) indicates how much additional information, measured in bits, the adversary needs to reconstruct the true values. This amount of information is similar to the amount of surprise the adversary will experience upon learning the true value, i.e. the information surprisal metric (see Figure \ref{fig:alz_information_surprisal} in the appendix).

Based on these findings in this and the previous sections, we rated each metric based on how easy it is to understand what its values mean, and how easily it can be interpreted.
We summarize the ratings in Table \ref{tab:privacymetrics} (column \textit{Intuitiveness}). 

\subsubsection{Interpreting the Overall Privacy Level}
The violin plots presented in Figure \ref{fig:alzheimer} are comprehensive, but they make it hard to tell what an individual's overall privacy level is against a specific adversary.
We propose radar plots to visualize the overall privacy level indicated by a combination of metrics.

\begin{figure}
\centering
 \subfloat[Uniform adversary]{\label{fig:radar_uniform}{\includegraphics[width=.32\linewidth]{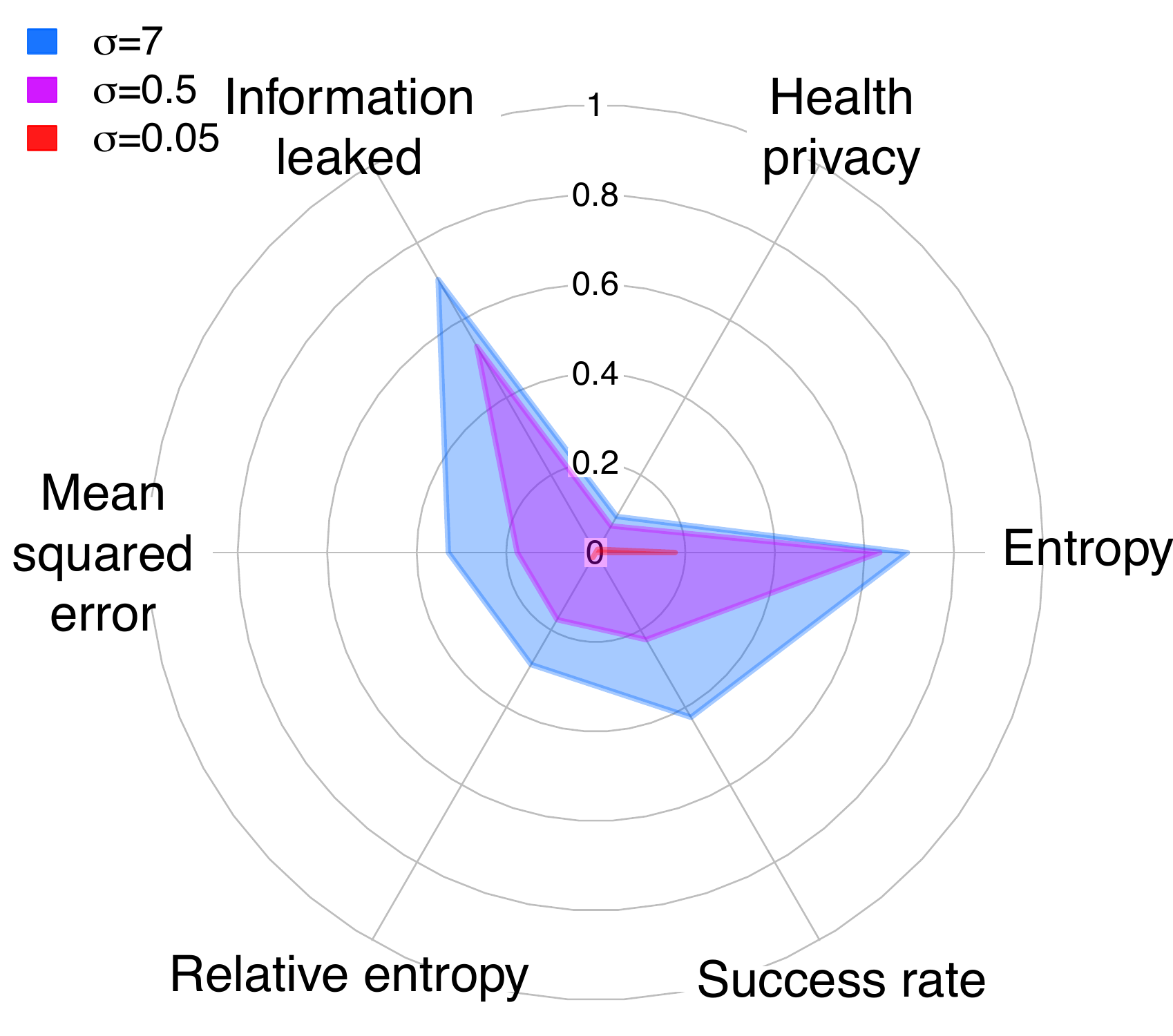} }}
 \subfloat[Normal adversary]{\label{fig:radar_normal}{\includegraphics[width=.32\linewidth]{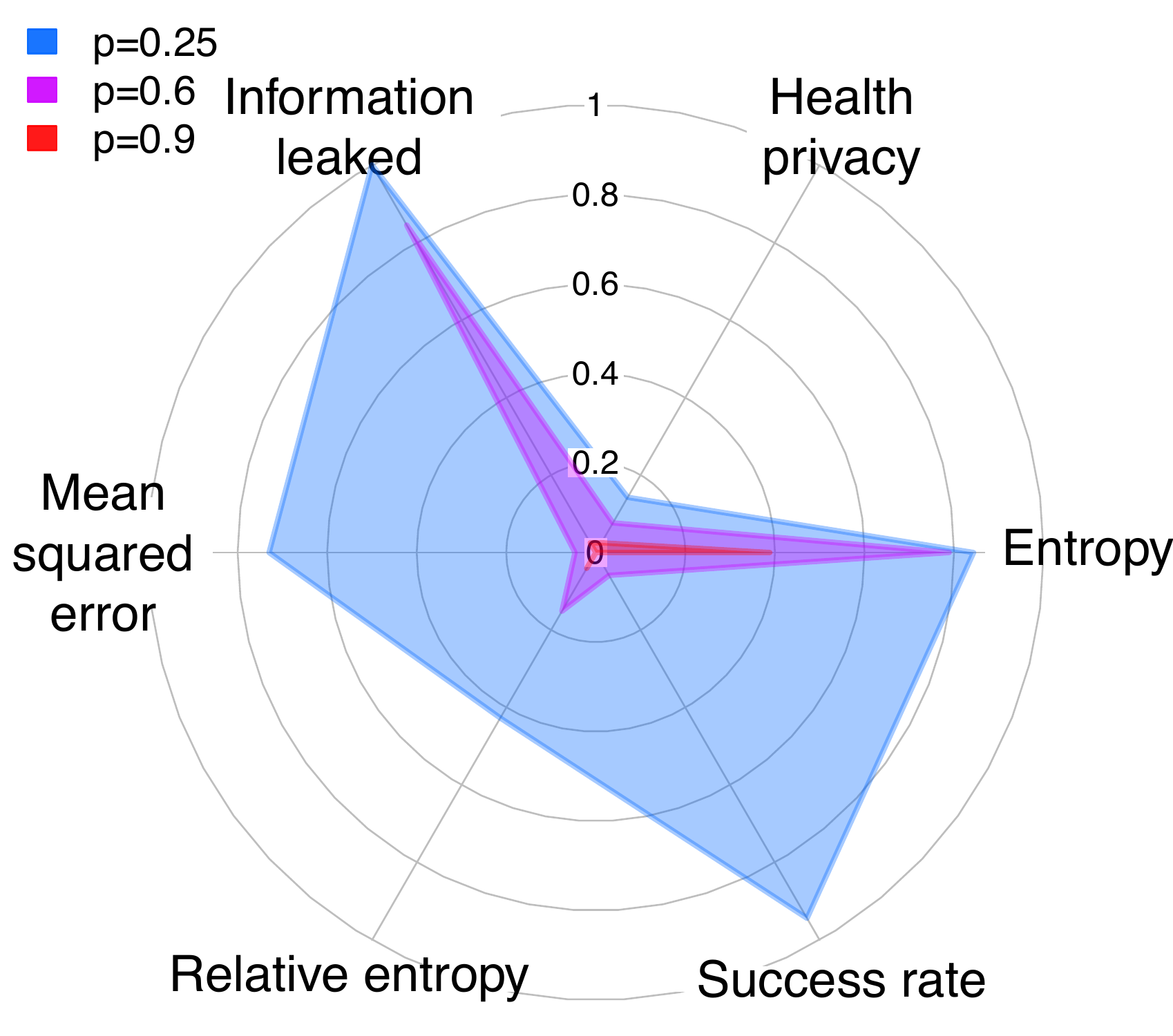} }}
 \subfloat[Reference adversary]{\label{fig:radar_reference}{\includegraphics[width=.32\linewidth]{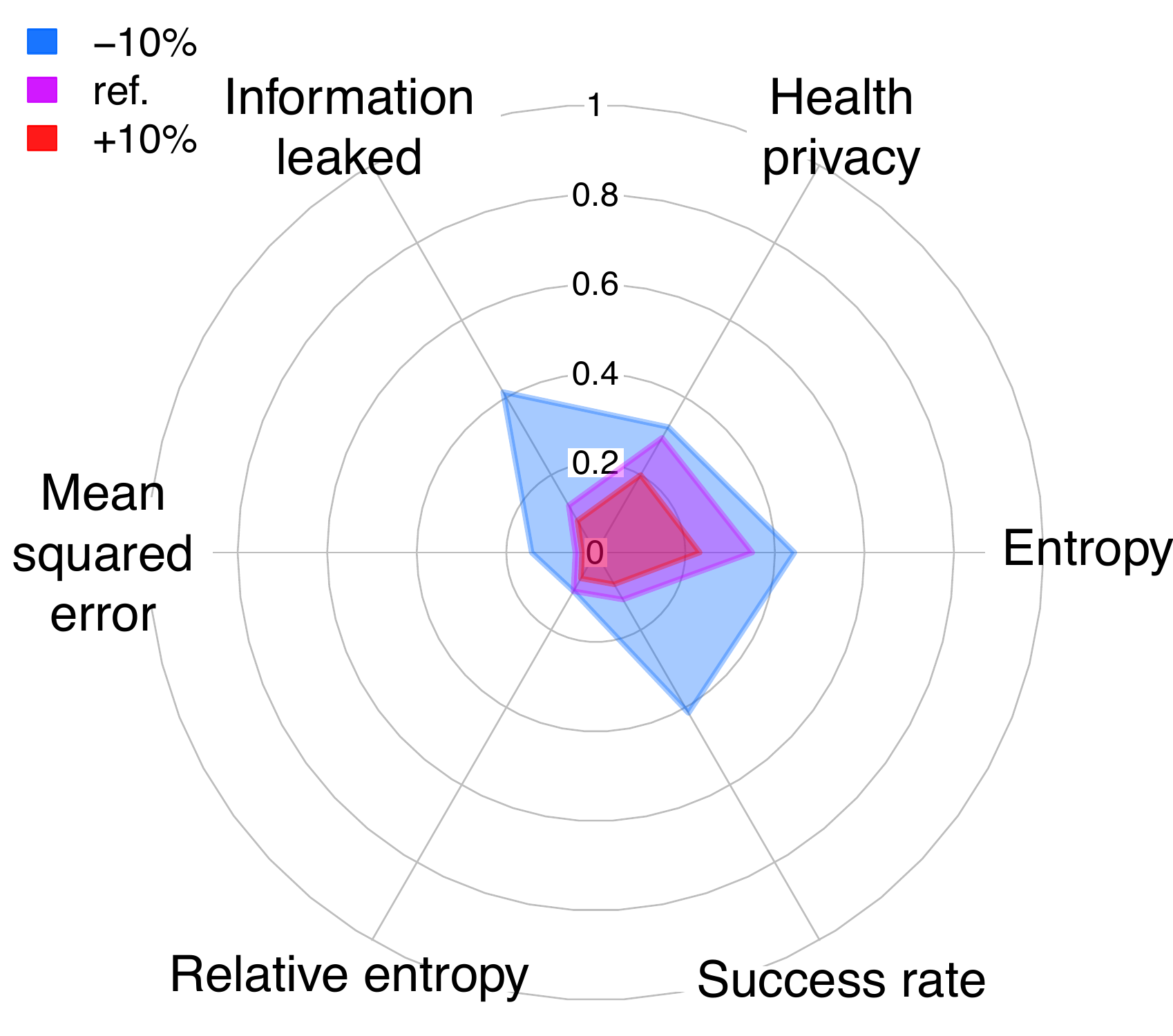} }}
  \caption{Radar plots visualizing the privacy level of six privacy metrics for three strength levels of each adversary type}%
  \label{fig:radar}%
\end{figure}

Figure \ref{fig:radar} shows one radar plot for each adversary type.
To keep the plots clean, we plot only three strength levels per adversary type. 
We excluded the expected estimation error because of its similarity to health privacy with the expected estimation error as base metric.
The values for each metric have been normalized to the $[0,1]$ value range using the 10th and 90th percentile of values across all adversary strengths. 
In addition, we inverted the values for lower-better metrics.
As a result, a larger area in the plots directly corresponds to a higher privacy level.

The plots allow comparisons between the different adversary types, for example clearly showing that the weakest \textit{normal} adversary (Figure \ref{fig:radar_normal}, light blue color) is much weaker than the weakest \textit{uniform} adversary in Figure \ref{fig:radar_uniform}, because the privacy area for the \textit{normal} adversary is much larger.
The plots also confirm our expectation that the reference adversaries are comparatively strong because their privacy areas are smaller than for the other two adversary types.
A real-world evaluation of privacy may not vary the adversary strengths as we have done, but instead vary parameters of a new privacy-enhancing technology.
Since the strength of the adversary and the strength of a \acl{PET} are essentially two sides of the same coin, we expect that radar plots will be able to highlight differences between \aclp{PET} in the same way as differences between adversaries.


\section{Conclusions and Future Work}

We measured the strengths of 23 published genomic privacy metrics.
We introduced monotonicity as the key indicator of a metric's strength, i.e. metrics should show decreasing privacy for increasing adversary strength.
We tested each of the 23 metrics in three different scenarios, for adversaries of different strengths, and found that only 7 out of 23 metrics were strong across scenarios and adversary types.
The 7 strong metrics were the adversary’s success rate, the amount of information leaked, health privacy (with information surprisal or relative entropy as base metric), information surprisal, percentage incorrectly classified, relative entropy, and user-specified innocence.
Furthermore, we found that none of the metrics we tested were sufficiently reliable when used in isolation. Therefore, we recommend that several strong metrics that measure different outputs should be used together.
Finally, we introduced two visualization methods previously not used in the privacy field -- heat maps and radar plots.
%
Our systematic comparison of genomic privacy metrics will enable researchers to make informed and consistent decisions about the selection of privacy metrics and \aclp{PET}.


In future work, we will measure the strength of privacy metrics in other application domains, e.g., vehicular networking and smart metering.
Future work also needs to study whether there are additional requirements for privacy metrics aside from monotonicity, and whether privacy metrics should satisfy the conditions for metrics in a mathematical sense.

\begin{acks}
This work used the ARCHER UK National Supercomputing Service (\url{http://www.archer.ac.uk}).
\end{acks}

\bibliographystyle{ACM-Reference-Format-Journals}
\bibliography{genomics-metrics}

\appendix
\section*{APPENDIX}
%

\section{Supplementary Figures}
\begin{figure}[h!]
 \centering
 \subfloat[Health Privacy (base: information surprisal)]{\label{fig:alz_health_privacy}{\includegraphics[width=\linewidth]{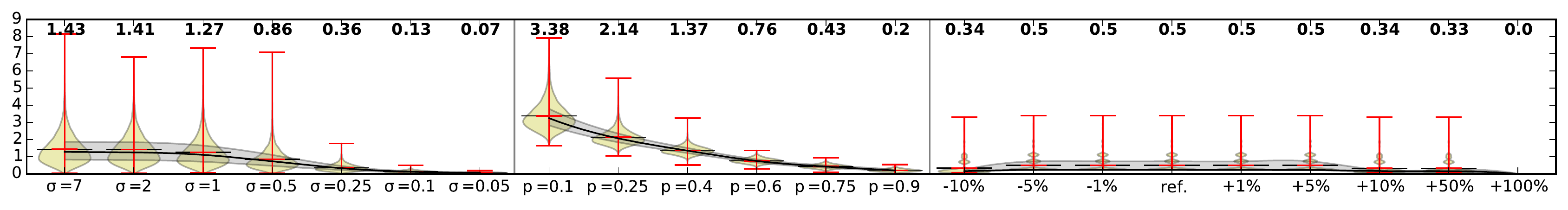} }}\\%
 \vspace{-.45cm}
 \subfloat[Information Surprisal]{\label{fig:alz_information_surprisal}{\includegraphics[width=\linewidth]{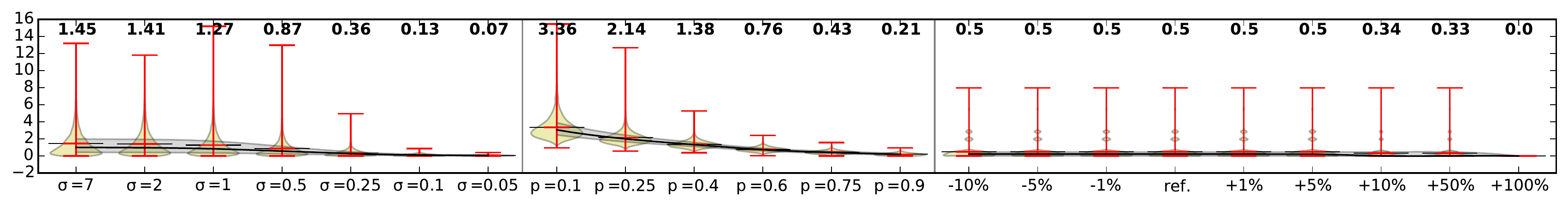} }}\\
 \vspace{-.45cm}
 \subfloat[Mean Error]{\label{fig:alz_mean_error}{\includegraphics[width=\linewidth]{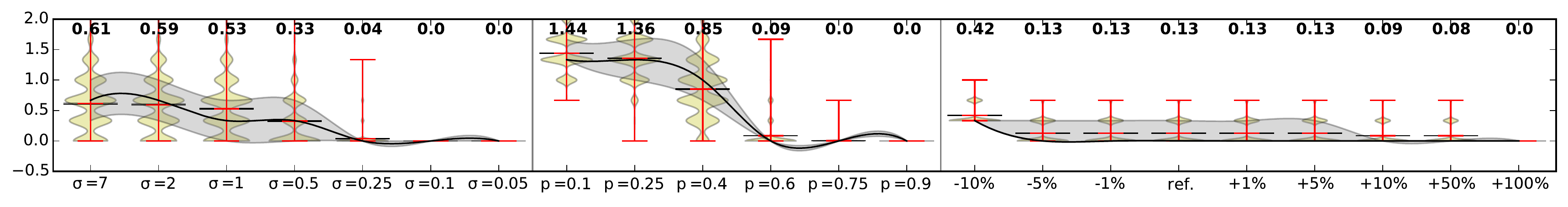} }}\\
 \vspace{-.45cm}
  \subfloat[Percentage Incorrectly Classified]{\label{fig:alz_percentage_incorrectly_classified}{\includegraphics[width=\linewidth]{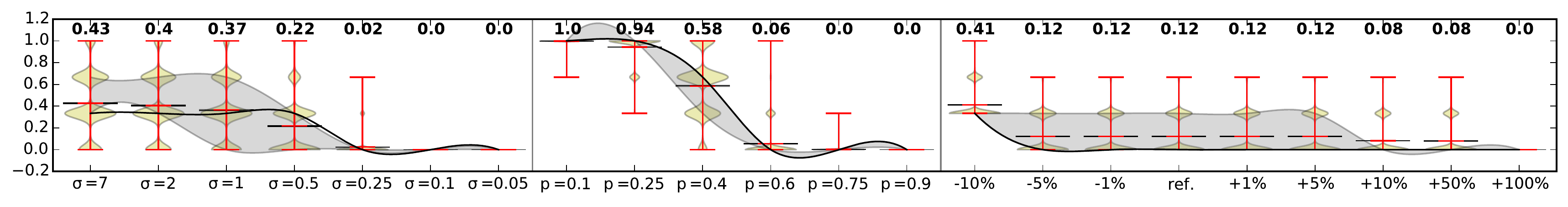} }}\\%
 \vspace{-.45cm}
 \subfloat[User-specified Innocence]{\label{fig:alz_user_specified_innocence}{\includegraphics[width=\linewidth]{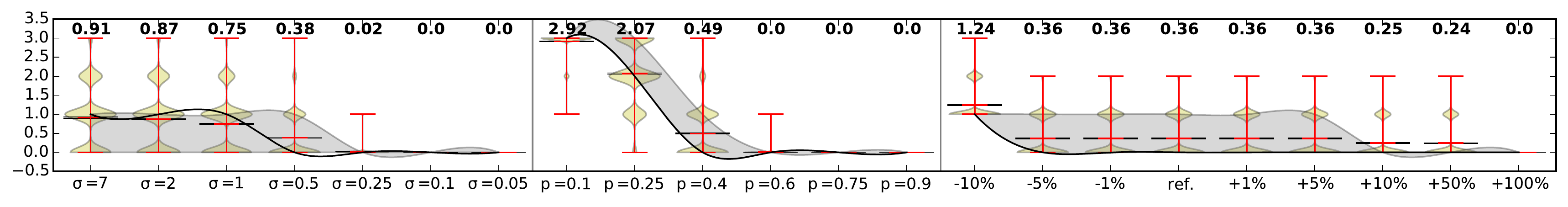} }}\\%
 \vspace{-.1cm}
 \caption{Strong privacy metrics for the Alzheimer's disease scenario, evaluated according to three adversary models. Adversary strengths for each model are ordered weakest to strongest from left to right.}%
  \label{fig:alzheimer_strong}%
\end{figure}

\begin{figure}[h!]
 \centering
 \vspace{-.45cm}
 \subfloat[Asymmetric Entropy (per SNP)]{\label{fig:alz_asymmetric_entropy_per_snp}{\includegraphics[width=\linewidth]{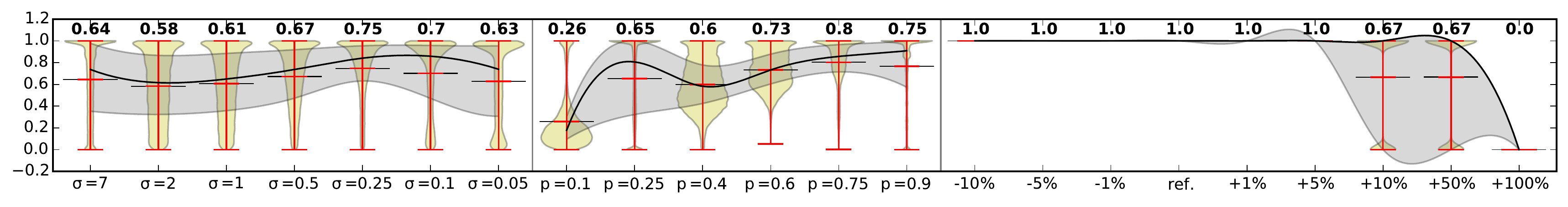} }}\\
 \vspace{-.45cm}
 \subfloat[Conditional Entropy]{\label{fig:alz_conditional_entropy}{\includegraphics[width=\linewidth]{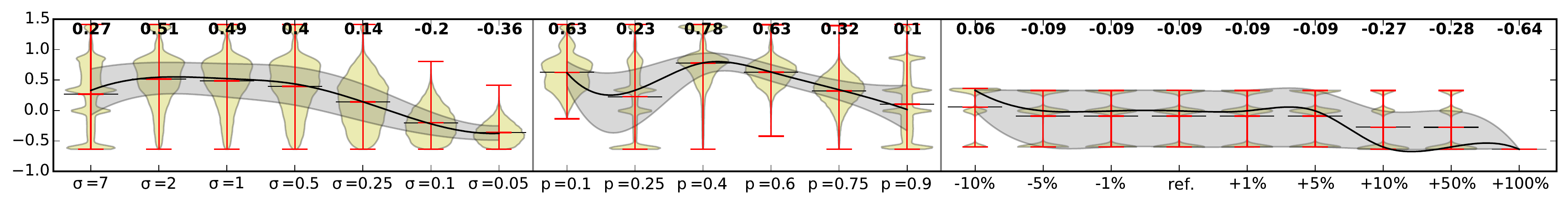} }}\\
 \vspace{-.45cm}
 \subfloat[Conditional Privacy Loss]{\label{fig:alz_conditional_privacy_loss}{\includegraphics[width=\linewidth]{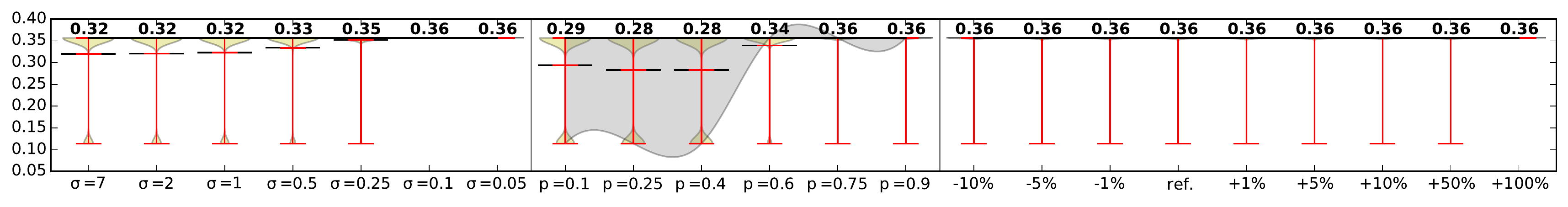} }}\\
 \vspace{-.45cm}
 \subfloat[Cumulative Entropy]{\label{fig:alz_cumulative_entropy}{\includegraphics[width=\linewidth]{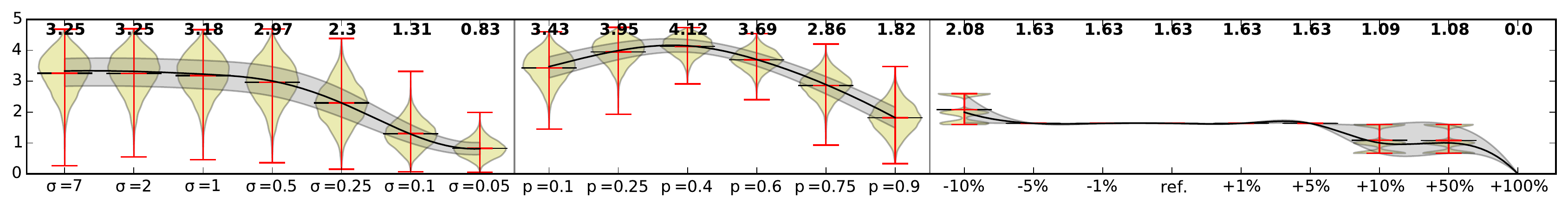} }}\\
 \vspace{-.45cm}
 \subfloat[Inherent Privacy]{\label{fig:alz_inherent_privacy}{\includegraphics[width=\linewidth]{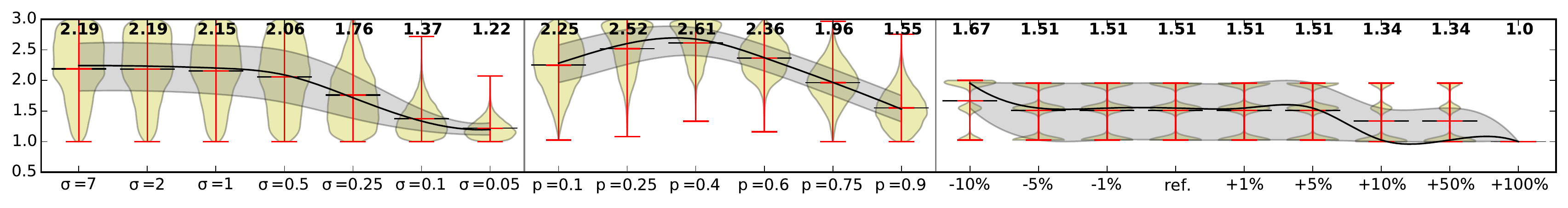} }}\\
 \vspace{-.45cm}
  \subfloat[Min-Entropy]{\label{fig:alz_min_entropy}{\includegraphics[width=\linewidth]{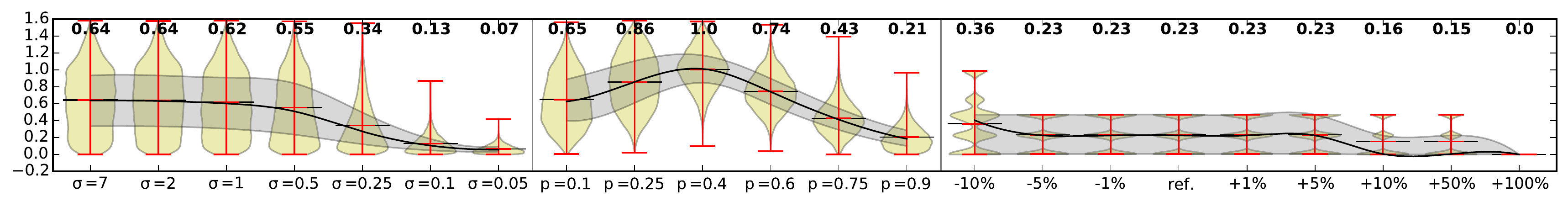} }}\\
 \vspace{-.45cm}
 \subfloat[Mutual Information]{\label{fig:alz_mutual_information}{\includegraphics[width=\linewidth]{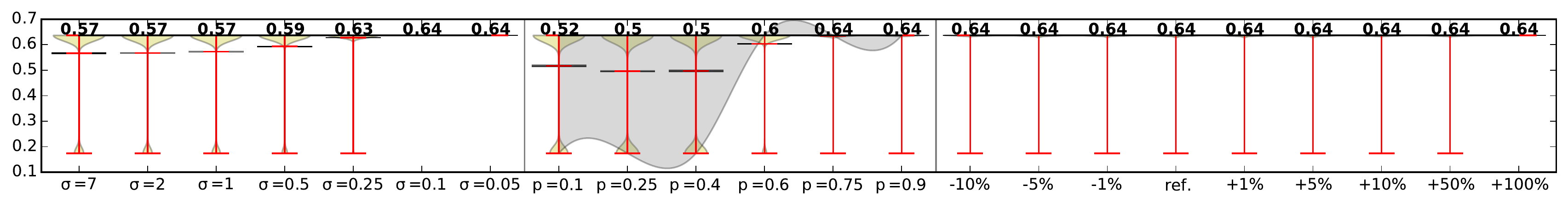} }}\\
 \vspace{-.45cm}
  \subfloat[Normalized Entropy]{\label{fig:alz_normalized_entropy}{\includegraphics[width=\linewidth]{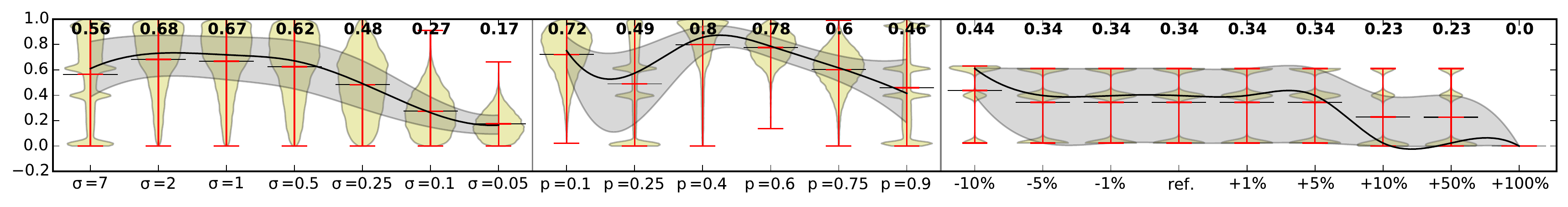} }}\\
 \vspace{-.45cm}
  \subfloat[Normalized Mutual Information]{\label{fig:alz_normalized_mutual_information}{\includegraphics[width=\linewidth]{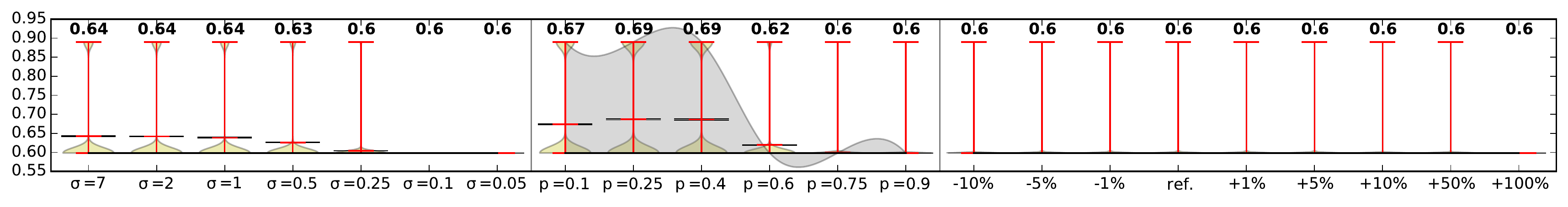} }}\\
 \caption{Average strength privacy metrics for the Alzheimer's disease scenario, evaluated according to adversary strength, ordered weakest to strongest from left to right}%
  \label{fig:alzheimer_average}%
\end{figure}

\begin{figure}[t]
 \centering
 \subfloat[Asymmetric Entropy]{\label{fig:alz_asymmetric_entropy}{\includegraphics[width=\linewidth]{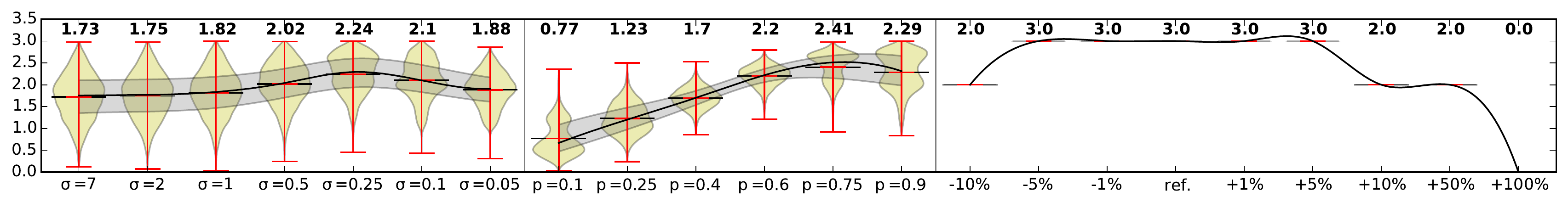} }}\\
 \vspace{-.45cm}
 \subfloat[Coefficient of Determination $r^2$]{\label{fig:alz_coefficient_of_determination}{\includegraphics[width=\linewidth]{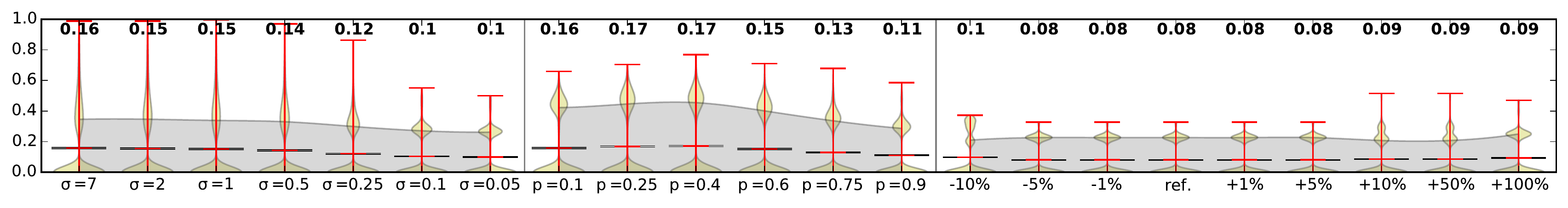} }}\\
 \vspace{-.45cm}
  \subfloat[Variation of Information]{\label{fig:alz_variation_of_information}{\includegraphics[width=\linewidth]{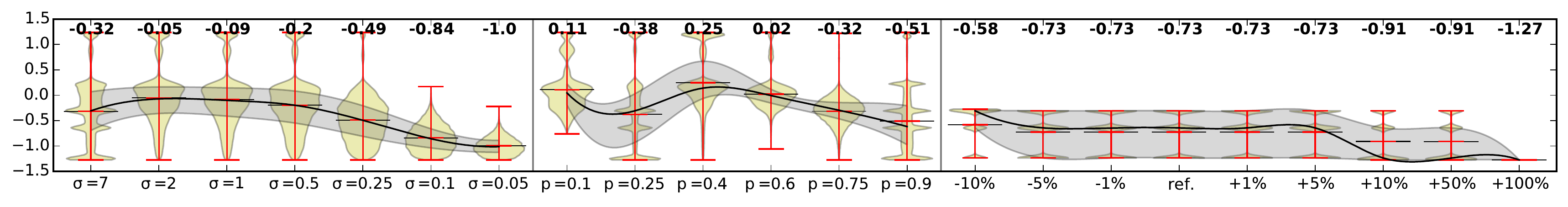} }}\\
 \caption{Weak privacy metrics for the Alzheimer's disease scenario, evaluated according to adversary strength, ordered weakest to strongest from left to right}%
  \label{fig:alzheimer_weak}%
\end{figure}

Figures \ref{fig:alzheimer_strong}, \ref{fig:alzheimer_average} and \ref{fig:alzheimer_weak} show strong, average strength, and weak privacy metrics for the alzheimer scenario which were not selected in Section \ref{sec:alzheimer}.

\end{document}